\documentclass[mathleft
]{an}
\usepackage{graphicx}
\usepackage{times}
\usepackage{longtable}
\usepackage{supertabular}
\usepackage{wasysym}
\usepackage{natbib}             
\bibpunct{(}{)}{;}{a}{}{,}      
\newcommand{\nm}[1]{_{\mbox{\scriptsize #1}}}
\newcommand{\halpha}{H$\alpha$}
\newcommand{\cahk}{Ca\,{\sc ii} H and K\ }
\newcommand{\kms}{km\,s$^{-1}$}
\newcommand{\MV}{$M_\mathrm{V}$}
\newcommand{\vsini}{\ensuremath{v\sin i}\ }
\newcommand{\Teff}{\ensuremath{T_\mathrm{eff}}}
\newcommand{\Porb}{\ensuremath{P_\mathrm{orb}}}
\newcommand{\Prot}{\ensuremath{P_\mathrm{rot}}}
\newcommand{\Msun}{\ensuremath{\mbox{M}_\mathrm{\odot}}}
\newcommand{\Rsun}{\ensuremath{\mbox{R}_\mathrm{\odot}}}
\newcommand{\Lsun}{\ensuremath{\mbox{L}_\mathrm{\odot}}}
\newcommand{\type}[2]{#1\,{\sc #2}}

\begin{document}

\Pagespan{1}{}
\Yearpublication{2006}%
\Yearsubmission{2005}%
\Month{11}%
\Volume{999}%
\Issue{88}%

\title{The chromospherically active binary star EI~Eridani}
\subtitle{I. Absolute dimensions} 

\author{A. Washuettl\inst{1,2}\thanks{Visiting Astronomer, Kitt Peak National Observatory and
      National Solar Observatories, National Optical Astronomy Observatory, which is operated
      by the Association of Universities for Research in Astronomy, Inc. (AURA) under cooperative
      agreement with the National Science Foundation.}
\and K. G. Strassmeier\inst{1}$^\star$
\and T. Granzer\inst{1}
\and M. Weber\inst{1}$^\star$
\and K. Ol\'ah\inst{2}
}
\institute{
        Astrophysical~Institute~Potsdam~(AIP),~An~der~Sternwarte~16,~D-14482~Potsdam,~Germany\\
        e-mail: awashuettl\,/\,kstrassmeier\,/\,tgranzer\,/\,mweber@aip.de
\and    Konkoly Observatory, P.O. Box 67, H-1525 Budapest, Hungary\\
        olah@konkoly.hu
}
\titlerunning{The active binary star EI~Eridani: absolute dimensions}
\authorrunning{A. Washuettl et al.}

\received{} \accepted{} \publonline{}

\keywords{stars: activity --
stars: individual: EI~Eri -- stars: late-type -- binaries: close} 

\abstract{We present a detailed determination of the astrophysical
parameters of the chromospherically active binary star
EI\,Eridani. Our new radial velocities allow to improve the set of
orbital elements and reveal long-term variations of the
barycentric velocity. A possible third-body orbit with a period of
$\approx 19$\,years is presented. Absolute parameters are
determined in combination with the {\em Hipparcos} parallax.
EI\,Eri's inclination angle of the rotational axis is confined to
$56\fdg0 \pm 4\fdg5$, its luminosity class {\sc iv} is confirmed
by its radius of $2.37 \pm 0.12$\,\Rsun. A comparison to
theoretical stellar evolutionary tracks suggests a mass of $1.09
\pm 0.05$\,\Msun\ and an age of $\approx6.15$~Gyr. The
present investigation is the basis of our long-term
Doppler imaging study of its stellar surface. }

\maketitle

\section{Introduction}
EI\,Eridani = HD 26337 (G5\,{\sc iv}, $P\nm{rot}\ = 1.947$ days,
$V$ = 7.1 mag) is a rapidly rotating ($v\sin i = 51$ \kms) active,
non-eclipsing, single-lined spectroscopic binary star and as such
a typical RS CVn star. RS CVn stars are close but detached binary
systems that rotate synchronously due to tidal forces. This
anomalously fast rotation is believed to be responsible for the
high activity level found on these stars and constitute valuable
laboratories to study the evolved Sun in action.

\cahk emission -- the classical fingerprint of magnetic activity -- on EI\,Eri was first noted
by \citet{bidelman73} and confirmed later by \citet{fekel80} who classified it as moderate in
strength, class C on the qualitative emission scale by \citet{hearn79}.
Line-profile data of \cahk and \halpha\ emission lines were obtained by
\citet{kgs:fekel90} and \citet{ff:montes94}.
\citet{fekel:hall82} classified EI~Eri as a single-lined RS~CVn star when they detected its
light variability, with an amplitude being almost 0\fm2 in $V$ and a photometric
and orbital period of around two days.
Later, \citet{fekel:moffett86} determined the orbital period to be
1\fd9472 while \citet{hall:osborn87} detected a photometric period
of $1.945 \pm 0.005$ days from $UBV$ photometry. The \halpha\ line
is in absorption, as in most RS\,CVn stars, but highly filled in
with chromospheric emission \citep[see][]{smith:bopp82} and quite
variable in strength \citep{fekel:moffett86,zboril:oliveira05}.
These variations are assumed to be caused by rotational modulation
of the chromospheric \halpha\ emission, presumably due to evolving
plage regions on the stellar surface moving in and out of sight.
\citet{fekel:quigley87}, \citet{pallavicini:randich92} and
\citet{randich:gratton93} reported the presence of moderate amounts
of lithium.
\citet{hall:osborn87} already noted season-to-season changes in
the photometric period ($\sim$1\%), in the light-curve amplitude
(0\fm07 - 0\fm20) and in the mean brightness ($\sim$10\%), likely
indicating latitude and/or longitude changes of the location of
starspots. \citet{kgs:hall89} report seasonal changes of the
photometric period of 0.043\,days within three seasons.
\citet{rodono:cutispoto92} suggested a possible cycle period of
about 10 years. \citet{kgs:bartus97} summarized the first 16 years
of photometric data and found a possible $11 \pm 1$ years cycle of
the mean brightness while \citep{olah:kgs02} refined this to be
12.2 years.
Recent observations in other spectral regions were presented by
\citet{osten:brown02}, \citet{garcia:paredes03} and \citet{cardini05}.

With its large rotational velocity and an intermediate inclination, EI\,Eri
is an ideal candidate for {\em Doppler imaging\/}.  This technique
\citep[see][]{rice02} allows the reconstruction of the surface spot distribution
of rapidly rotating stars by using the relation between wavelength position
across an absorption line and spatial position across the stellar disk.
However, the rotational period of 1.947 days makes it difficult to obtain
spectral coverage for a complete rotation period, and ideally, when
using a single observing site, three weeks of observations are needed
to provide good phase coverage for a Doppler image of EI\,Eridani.

Doppler images for the 1984-87 period were presented by
\citet{hatzes:vogt92}. \citet{kgs90} and \citet{kgs:rice91}
published Doppler images from 1987-88. All images show a large
asymmetric spot at the pole or at high latitudes with several
appendages and occasionally equatorial spots. The polar spot is
long lived but exhibits significant changes of its size and shape,
while the equatorial spots change within weeks. Furthermore,
\citet{donati:semel97} detected a clear Zeeman signature of the
magnetic field of the primary star.

In this paper, we present spectroscopic and photometric
observations obtained at KPNO and NSO in the years 1988 -- 1998
and with the MUSICOS~98 campaign, and perform an investigation of
the absolute dimensions of EI\,Eri in order to prepare for a
Doppler imaging study. In two forthcoming paper, we will 
present our Doppler imaging results and address the question of spot
lifetimes, long-term cyclic behavior \citep[][ hereafter ``paper {\sc ii}'']{paper2} and
differential rotation \citep[][ hereafter ``paper {\sc iii}'']{paper3}.

\section{Observations and data reduction}
All our spectroscopic data were obtained from four observing runs with
the Coud\'e Feed telescope at Kitt Peak (March 1994, February 1995, January 1996,
December 1997), one dedicated visitor observing run
at the McMath-Pierce Telescope (NSO) in Nov/Dec. 1996, from the McMath/NSO synoptic
program (1988--1995) and from the MUSICOS campaign 1998.
The observing logs and radial velocities from KPNO and NSO are given in Table~\ref{tab:obslog}
which is available only in electronic form.
The MUSICOS\,98 spectroscopic data will be presented in a forthcoming dedicated paper.
A summary of all observing runs is shown in Table~\ref{tab:obsfac}.

\begin{table*}[htbp]
\caption{Overview of all observing runs.}\label{tab:obsfac}
\begin{center}
\begin{tabular}{lcc@{, }l@{ -- }c@{, }lcrclcc}
\hline
\noalign{\smallskip}
Site    & Acronym & \multicolumn{4}{c}{Date} & HJD   & n\,$^a$ & Detector & Disp.    & ThAr & $\lambda/\Delta\lambda$ \\
        &         & \multicolumn{4}{c}{}     & (24+) &         &          & [\AA/px] & FWHM & [\AA]            \\
        &         & \multicolumn{4}{c}{}     &       &         &          &          & [px] &                  \\
\noalign{\smallskip}
\hline
\noalign{\smallskip}
NSO     & NSOsyn  & 1988 & Nov 16 & 1995 & Dec 28 & 47481.9 -- 50079.9 &225 & TI-4 & 0.096 & 1.6 & 42\,000 \\
        & M96     & 1996 & Nov 01 & 1996 & Jan 08 & 50388.9 -- 50457.8 & 58 & TI-4 & 0.118 & 1.9 & 29\,000 \\
KPNO    & CF94    & 1994 & Mar 05 & 1994 & Mar 15 & 49416.6 -- 49426.6 &  3 & TI-5 & 0.106 & 1.7 & 36\,000 \\
        & CF95    & 1995 & Feb 21 & 1995 & Mar 06 & 49770.6 -- 49783.6 &  5 & TI-5 & 0.106 & 1.7 & 36\,000 \\
        & CF96    & 1996 & Jan 10 & 1996 & Jan 24 & 50093.6 -- 50107.8 & 24 & F3KB & 0.104 & 2.6 & 24\,000 \\
        & CF97    & 1997 & Dec 27 & 1997 & Jan 15 & 50809.6 -- 50828.9 & 57 & TI-5 & 0.1055& 2.0 & 30\,000 \\
\multicolumn{2}{l}{MUSICOS\,$^b$} & 1998 & Nov 21 & 1998 & Dec 13 & 51139.4 -- 51159.6 & 95 & {\em (various)} &    &     &         \\

\noalign{\smallskip}
\hline
\noalign{\smallskip}
\multicolumn{10}{l}{$^a$ \dots\ Number of spectra}\\
\multicolumn{10}{l}{$^b$ \dots\ The results from the MUSICOS campaign will be presented in a dedicated paper.}\\
\end{tabular}
\end{center}
\end{table*}

\subsection{Kitt Peak National Observatory (KPNO)}
The KPNO data were obtained with the 0.9m Coud\'e feed telescope.
For the 1994 run (March 4 -- 18; hereafter:
``CF94'') and the 1995 run (Feb. 22 -- March 7; ``CF95'') we used the
800 $\times$ 800 TI-5 CCD chip (15$\mu$m pixels), grating A, camera 5, the long
collimator and a 280\,$\mu$m slit giving a resolving power ($\lambda/\Delta\lambda$) of
36\,000 at a wavelength of 6420\,\AA\ (the FWHM of an unblended Th-Ar
comparison lamp line was about 1.7 pixel; dispersion was 0.1\,\AA/px).
The useful wavelength range of the resulting spectra is 80\,\AA.
For the 1996 run (Jan. 11 -- 25; ``CF96'') we used the 1000 $\times\ $ 3000 pixel Ford CCD
(F3KB chip, $15 \mu$m pixels; two pixels each were binned), with an otherwise
identical spectrograph setup.
F3KB enables a much larger wavelength region ($\sim$320\,\AA\ from $\lambda$6335 to
$\lambda$6655) at a
spectral dispersion of $\approx$ 0.104\,\AA\ per pixel. The resolving power is 24\,000
at 6420\,\AA\ as measured by the FWHM of a Th-Ar line of about 2.6 pixel.
An additional observing run primarily dedicated to EI\,Eridani was undertaken in
winter 1997/98 (Dec. 26 -- Jan. 15; ``CF97'').
This time, we employed the TI-5 chip again, seeking the better resolution.
The FWHM of a Th-Ar line was measured to be 2.0 pixel, corresponding to a resolving power
of approximately 30\,000.

A typical night of observations includes 20--40 biases, 10--20 flat-field
images, and 2--3 integrations of a thorium-argon hollow cathode lamp for
establishing the wavelength scale, at least one at the beginning and one at the
end of each night, but usually every three hours.
Additionally, at least one observation from a choice of four radial velocity standard
stars (see Table~\ref{tab:rvss}) was taken per night.
Neither the TI-5 CCD nor the F3KB CCD show discernible signs of fringing
at the wavelength of our observations. Consequently, no attempts were made to
correct for it other than the standard flat-field division.

Data reduction was done using the NOAO/IRAF software package and followed
our standard procedure as described in \citet{paperv} which includes
bias subtraction, flat fielding, and optimal aperture extraction.
All spectra were checked for narrow telluric water lines. These
lines can be blended with Doppler imaging lines and produce
misleading artifacts in the surface maps. The spectra of a rapidly
rotating ($v \sin i \approx 300$ \kms) hot B-star revealed these
telluric lines and were compared with the spectra of EI\,Eri.
Sporadic telluric water-vapor lines were found in this wavelength region, but
fortunately, in most cases no blend of a water line
with a Doppler imaging line occurred, and no corrections had to be
applied to the line profiles used for further analysis.
Special care was exercised during the continuum fitting process.
A low-order polynomial was sufficient to find a satisfactory continuum solution.
Exposure length was typically 45 minutes but up to 60 minutes in case of
thin cirrus.

With this integration time we achieved a signal-to-noise ratio of up to 400:1.

\begin{table}
\begin{center}
\caption{Radial velocity standard stars. The velocity values are from \citet{scarfe90}
except 59~Ari which is from \citet{wilson53}.}
\label{tab:rvss}
\begin{tabular}{@{}r|c@{\,\,\,\,\,}c@{\,\,\,\,\,}c@{\,\,\,\,\,}c@{\,\,\,\,\,}c@{\,\,\,\,\,}c@{}}
\hline
\noalign{\smallskip}
        & $\alpha$\,Tau  & $\alpha$\,Ari  & $\beta$\,Gem   & 16\,Vir  & 59\,Ari & $\beta$\,Lep \\
\noalign{\smallskip}
\hline
\noalign{\smallskip}
HD      & 29139         & 12929         & 62509         & 107328  & 20618  & 36079 \\
$v_r$   & 54.25         & -14.51        & 3.23          & 36.48   & -0.1   & -13.5 \\
$\sigma_{v_r}$ & 0.08   & 0.11          & 0.15          & 0.04    & ?      &  0.1  \\
\noalign{\smallskip}
\hline
\noalign{\smallskip}
 NSOsyn &               &               &               &         & *      &   \\
 NSO 96 &               &       *       &               &         &        &   \\
  KP 94 &               &               &               &       * &        &   \\
     95 &       *       &               &               &       * &        &   \\
     96 &       *       &               &       *       &       * &        &   \\
     97 &               &       *       &       *       &       * &        &   \\
{\footnotesize MUSICOS} & * &           &               &         &        & * \\
\noalign{\smallskip}
\hline
\end{tabular}
\end{center}
\end{table}

\begin{table}
\caption[Original and improved orbital elements]
{Original and improved orbital elements for EI\,Eri.
The original values are from \citet{kgs90} except $T_0$ which is from
\citet{fekel:quigley87} and $M_1$ from \citet{nordstrom:mayor04}.
The numbers in brackets denote the error in the last digit(s).}\label{tab:orbit}
\begin{center}
 \begin{tabular}{lll}
  \hline
  \hline
  \noalign{\smallskip}
  Orbital element      & Original         & Improved \\
  \noalign{\smallskip}
  \hline
  \noalign{\smallskip}
  $P$\ [days]       & 1.947227 (8)        & 1.9472324 (38)       \\
  $T_0$ [HJD]       & 2\,446\,074.384     & 2\,448\,054.7109 (5) \\
  $\gamma$ [\kms]   & 17.6 (2)            & 21.64 (16)           \\
  $K_1$ [\kms]      & 27.4 (3)            & 26.83 (23)           \\
  $e$ (assumed)     & 0.0                 & 0.0                  \\
  $a_1 \sin i$ [km] & 733000 (9000)       & 718400 (6200)        \\
  f\,(M) [\Msun]    & 0.00415 (15)        & 0.00391 (10)         \\
  \noalign{\smallskip}
  \hline
  \noalign{\smallskip}
  $M_1$ [\Msun] $^a$     & 1.10 (12)       & 1.09 (5)            \\
  $M_2$ [\Msun] $^b$     & ~~--            & 0.23 (4)            \\
  $K_2$ [\kms] $^c$      & ~~--            & 128 (25)            \\
  $a_2 \sin i$ [km] $^c$ & ~~--            & 3.4 (7) $\cdot 10^6$\\
  $a$ [km] $^d$          & ~~--            & 5.0 (9) $\cdot 10^6$\\
  \noalign{\smallskip}
  \hline
  \noalign{\smallskip}
  \multicolumn{3}{l}{Standard error of an observation} \\
  \multicolumn{2}{l}{\ \ \ of unit weight [\kms]} & 3.2 \\
  \noalign{\smallskip}
  \hline
  \hline
  \noalign{\smallskip}
  \multicolumn{3}{l}{\footnotesize $^a$ \dots\ estimated using evolutionary tracks; see \S\ref{sec:mass}.} \\
  \multicolumn{3}{l}{\footnotesize $^b$ \dots\ calculated using $f(\mbox{M})$, assuming $i = 56.0 \pm 4.5\degr$.} \\
  \multicolumn{3}{l}{\footnotesize $^c$ \dots\ using the estimations of $M_1$ and $M_2$.} \\
  \multicolumn{3}{l}{\footnotesize $^d$ \dots\ assuming $i = 56.0\pm 4.5\degr$.} \\
\end{tabular}
\end{center}
\end{table}

\begin{table}
\caption[Third body orbit]
{Suggestions for a possible third body orbit.
The index ``1-2'' refers to the inner binary system.}\label{tab:tertiary}
\begin{center}
 \begin{tabular}{lll}
  \hline
  \hline
  \noalign{\smallskip}
  Orbital element      & circular         & eccentric\\
  \noalign{\smallskip}
  \hline
  \noalign{\smallskip}
  $P$\ [days]           & 6700 $\pm$ 500          & 6700 $\pm$ 500          \\
  $T_0$ [HJD] (adopted) & 2\,442\,500             & 2\,443\,300             \\
  $\gamma$ [\kms]       & 21.0                    & 19.4                    \\
  $K_{1-2}$ [\kms]      & 4 $\pm$ 1               & 5.8 $\pm$ 1             \\
  $e$ (assumed)         & 0.0                     & 0.4                     \\
  $\omega$ [\degr]      & --                      & 60                      \\
  $a_{1-2} \sin i$ [km] & $3.7 \pm 1.0\cdot 10^8$ & $5.3 \pm 2.3\cdot 10^8$ \\
  f\,(M) [\Msun]        & 0.044 $\pm$ 0.035       & 0.13 $\pm$ 0.18         \\
  \noalign{\smallskip}
  \hline
  \noalign{\smallskip}
  $M_{1-2}$ [\Msun]      & $1.32 \pm 0.07$      & $1.32 \pm 0.07$       \\
  $M_3$ [\Msun] $^a$     & $0.7 \pm 0.6$        & $1.1 \pm 1.7$         \\
  $K_3$ [\kms] $^b$      & $8 \pm 7$            & $7 \pm 11$            \\
  $a_3 \sin i$ [km] $^b$ & $7 \pm 7 \cdot 10^8$ & $6 \pm 10 \cdot 10^8$ \\
  $a$ [AU] $^c$          & $9 \pm 6$            & $9 \pm 8$             \\
  \noalign{\smallskip}
  \hline
  \noalign{\smallskip}
  \multicolumn{3}{l}{Standard error of an observation of unit} \\
  \multicolumn{1}{l}{\ \ \ weight [\kms]} & 3.3 & 1.6 \\
  \noalign{\smallskip}
  \hline
  \hline
  \noalign{\smallskip}
  \multicolumn{3}{l}{\footnotesize $^a$ \dots\ calculated using $f(\mbox{M})$, assuming $i = 56.0 \pm 4.5\degr$.} \\
  \multicolumn{3}{l}{\footnotesize $^b$ \dots\ using the estimations of $M_{1-2}$ and $M_3$.} \\
  \multicolumn{3}{l}{\footnotesize $^c$ \dots\ assuming $i = 56.0\pm 4.5\degr$.} \\
\end{tabular}
\end{center}
\end{table}

\subsection{National Solar Observatory (NSO)}\label{chap:obs:nso}
Our NSO data were obtained with the 1.5\,m~McMath-Pierce solar
telescope during a regular visitor observing run covering 70
nights (from November 1, 1996, to January 8, 1997; hereafter:
``M96''), as well as from the synoptic nighttime program
\citep{smith:giampapa87} from 1988 till 1995. The 58 spectra from
the M96 observing run were taken using the 4.6m vertical
Czerny-Turner stellar spectrograph \citep{smith:jaksha84} with the
Milton-Roy grating \#1 and the 105-mm transfer lens together with
the 800$\times$800 TI-4 CCD camera (15$\mu$m pixels) at a
dispersion of 0.118 \AA/pixel. The mean FWHM of Thorium-Argon
emission lines was around 1.9 pixels, corresponding to a resolving
power of $\sim$29\,000. The spectra were generally recorded in a
sequence of 3$\times$15 minutes exposures and added up later
including cosmic ray removal. The integration time of the combined
spectra was between 20 and 60\,min. The observations covered a
wavelength range of about 45\,\AA\ from 6410 to 6460 \AA\ and
included three lines suitable for Doppler imaging: Fe\,{\sc
i}\,6411, Fe\,{\sc i}\,6430 and Ca\,{\sc i}\,6439. The average
signal-to-noise ratio is $\approx$150:1. Data reduction was done
again using the NOAO/IRAF software package in the same way as for
the KPNO coud\'e feed data. Usually, 20 flat-field exposures with
a Tungsten lamp were taken at the beginning and at the end of each
night, about three exposures of a Thorium-Argon reference lamp and
at least one observation of the IAU radial velocity standard star
$\alpha$~Ari were taken each night. The 58 spectra in M96 cover
together 35.4 consecutive stellar rotations.

The other part of the NSO observations were part of the synoptic
night-time program at the McMath-Pierce telescope during the years
1988 till 1995. Over the course of these seven years, a total of
225 spectra was acquired. The same 800$\times$800 TI-4 was applied
in conjunction with the Milton-Roy (B\&L) grating \#1
but with a significantly better dispersion of 0.096 \AA/px and an average FWHM of a comparison lamp
emission line of 1.6\,px, thus increasing the resolution to R=42\,000,
giving an effective wavelength resolution of 0.15\,\AA\ \citep[see][]{kgs:rice91}.
The signal-to-noise ratio of the NSO synoptic spectra reaches in most cases 300:1.

\subsection{Photometric observations}
Photometric data were provided by the Amadeus (T7) 0.75m Vienna-Observatory
Automatic Photoelectric Telescope (APT), part of the University of Vienna twin APT,
now at Washington Camp in southern Arizona \citep{kgs:boyd97}, previously,
before summer 1996, on Mt. Hopkins.
The observations were made differentially with respect to HD\,25852 as the
comparison star 
($V = 7\fm83$; since Nov. 8, 1996) 
and HD\,26409 as check star ($V = 5\fm448$).
All photometry was transformed to match the Johnson-Cousins $V$($RI$)$_C$ system
\citep[see][ \S4]{wooz:reegen01}.
In addition, older photometric observations back to 1980 were used from the literature
\citep[see][ Table~5]{kgs:bartus97}.

\section{Orbit analysis}

\subsection{Radial velocities and spectroscopic orbit}\label{chap:orbit}

Radial velocities (RVs) were determined from cross-correlations with spectra
of a radial velocity standard star observed during the same night.
Since EI\,Eri is a single-lined spectroscopic binary, only the RV
of the primary component can be measured. No RV information
is available for the secondary star.
The cross correlations were computed using IRAF's {\tt fxcor} routine
\citep[see][]{fitz93}, which fits a Gaussian to the cross-correlation
function to determine the central shift. Several different IAU standards
were used, see Table~\ref{tab:rvss}. Their RVs were adopted from the
work of \citet{scarfe90} except 59~Ari which is from \citet{wilson53}.

Orbital elements were derived with the differential-correction
program of \citet{barker67}, as modified and described by
\citet{fekel:kgs99}. We obtained altogether 354 new RVs for the
primary component of EI\,Eri, 89 of which are from our four
dedicated KPNO Coud\'e Feed observing runs, 58 from the one
dedicated visitor observing run at NSO/McMath (M96). Additionally,
178 from the NSO synoptic night-time program (1988--1995) and 29
from the MUSICOS\,98 campaign. Additionally, 55 older RVs were
found in the literature: 27 from \citet{fekel:quigley87}, 21 from
\citet{kgs90}, five from \citet{gunn:hall96}, and two from
\citet{donati:semel97}. Fekel \& Strassmeier used the RV standards
$\alpha$~Tau, $\beta$~Gem, 10~Tau and $\iota$~Psc with the RV
values from the older IAU list of RV standard stars
\citep{pearce55}. Therefore, they were corrected to match the
values by \citet{scarfe90}. The values by \citet{gunn:hall96} and
\citet{donati:semel97} were not used for calculating
the orbit. 

\begin{figure}
\includegraphics[width=83mm,clip,angle=0]{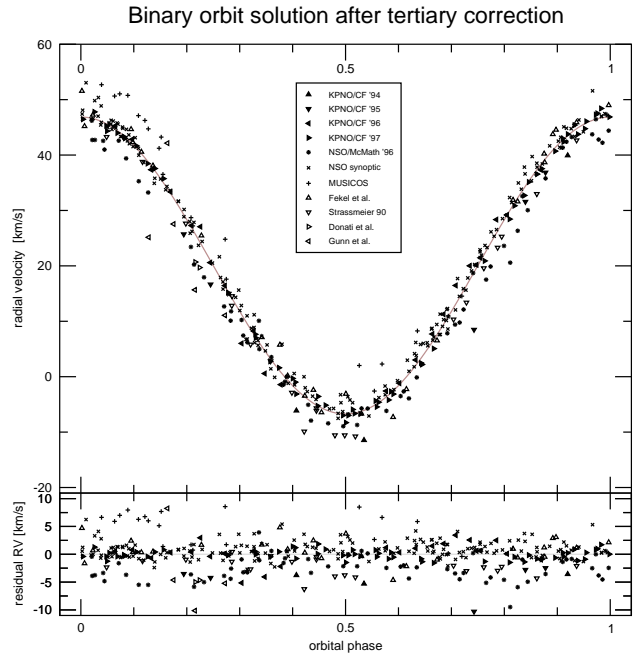}
\caption[]{Observed and computed radial velocity curve. The line is
the orbital solution from the elements in Table~\ref{tab:orbit}. The lower panel
gives the residual radial velocities, i.e. the difference between the observed
radial velocities and the theoretical curve from the orbital solution.
A third body in a wide orbit was taken into account.}
\label{orbit}
\end{figure}

Table~\ref{tab:orbit} lists the improved orbital elements from minimized {\em O$-$C\/} residuals
using all available RVs. The orbit is plotted in Fig.~\ref{orbit}.
Our final orbital elements converged at an eccentricity so close to zero that,
in accordance with the criteria established by \citet{lucy:sweeney71}, a formal
zero-eccentricity solution was adopted.
The standard error of an observation of unit weight is 3.2 \kms .
Phases of all line profiles were then computed using our revised ephemeris
\begin{center}
\begin{equation}
\mathrm{HJD} = 2,448,054.7109 + 1.9472324 \times\ E \ .
\end{equation}
\end{center}

Our revised orbital elements agree very well with the older ones from
\citet{fekel:quigley87}, except for $\gamma$ (see \S\ref{sec:triple}).\footnote
{\citet{donati:semel97} pointed out that the RVs they measured
at two epochs are in strong disagreement with the ephemeris from \citet{kgs90}.
\citet{gunn:hall96} also report that their five RVs
diverge significantly; 
they suspect the high rotational velocity to be responsible for the large
discrepancies.
As a matter of fact, Strassmeier's $T_0$ value for his orbital
solution seems to have been confused possibly due to a typing mistake
as it does not give sensible results for his own RVs.
Accordingly, both Donati's and Gunn's RV values are in perfect
agreement with the previous ephemeris of \citet{fekel:quigley87}.}

\subsection{EI\,Eri -- a triple-star system?}\label{sec:triple}
\begin{figure}
\includegraphics[width=65mm,angle=270,clip]{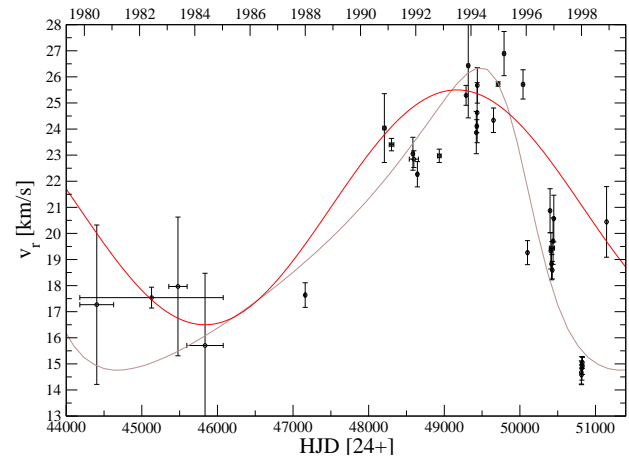}
\caption{Variations of the barycentric velocity $\gamma$.
The bars in the axis of abscissae (x) denote the time span of the data used for calculating the orbit.
The bars in the axis of ordinates (y) are the error bars of the respective $\gamma$ values.
The curve shows two suggestions for tertiary orbits according to the orbital
parameter in Table~\ref{tab:tertiary}.}
\label{gammavar}
\end{figure}

By obtaining precise radial velocity measurements,
more and more active close
binary stars turn out to be members of triple systems, e.g.
        BY\,Dra \citep{zuckerman:webb97},
        XY\,UMa \citep{chochol:etal98},
        HU\,Vir \citep{fekel:kgs99},
        UX\,Ari \citep{duemmler:aarum01}.
Also for EI\,Eri, we noticed that the barycentric velocity $\gamma$ seemed to have
increased since the early observations in the 1980s by up to 8\,\kms\ in the mid-1990ies.
In order to examine whether the system is possibly a triple star, we calculate an orbit for several
data subsets just varying the barycentric velocity $\gamma$ and, for comparison, also the amplitude
of the semi-major axis $K_1$ (Fig.~\ref{gammavar}).
The $\gamma$ values of EI\,Eri vary between 15.7 and 26.7\,\kms\ ($>3\sigma$)
with an average error of 0.8\,\kms.
It is seen that $\gamma$ shows an obvious trend over the period of 20 years, which is not the case
for $K_1$. A possible explanation could be the use of different RV standard stars
and the use of different instrumental configurations and reduction techniques.
The standards used for measuring the RV shifts for the KPNO, NSO and
MUSICOS observing runs are listed in Table~\ref{tab:rvss}.
\citet{kgs90} and \citet{fekel:quigley87} largely used the same standard stars.
For the observing runs by \citet{fekel:quigley87}, \citet{kgs90} and our CF97 run, two of the three
RV standards used are in common. However, $\gamma$ differs by 4\,\kms\ which is about
10$\sigma$. The NSO data, all taken with the same instrumental
setup and the same RV standard star (except M96) vary by 9\,\kms\ (mean deviation: 2--3\,\kms).
A cross correlation of all RV standard stars used (NSO, KPNO and MUSICOS)
uncovers a mean standard deviation of 0.82\,\kms\ for the RV standard stars of the individual
observing runs, while the mean RV values of the blocks exhibit a standard deviation of 1.69\,\kms\
(2.1\,$\sigma$). Only the observing blocks CF96 (3.2$\sigma$), CF97 (5.6$\sigma$) and MUSICOS
(3.6$\sigma$) deviate more than 2\,$\sigma$. During the same period, $\gamma$ varies
on average by 7\,$\sigma$. We conclude that the $\gamma$~variations are not due to stochastic
variations of the RV standard stars (as caused by, e.g., different instrumental setups).

\begin{figure}
\includegraphics[width=83mm,angle=0,clip]{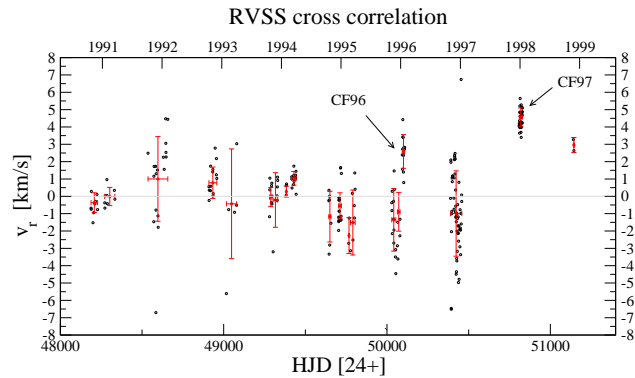}
\caption[Cross correlation of all radial velocity standard stars]
{Cross correlation of all used radial velocity standard star spectra.
The dots are the radial-velocity offsets for all individual RV standard stars (RVSS).
The dots with bars are the mean RV value for each observing block (corresponding to the suborbits
in Fig.~\ref{gammavar}); the bars in the axis of abscissae (x) denote the time span of the respective
observing block, the bars in the axis of ordinates (y) are the standard deviations of the RV values of
the respective observing block.}
\label{rvss_cross_correlation}
\end{figure}

As a test, we take the (observed-minus-computed) RV residuals of
the orbit solution of the primary-secondary system presented in
Fig.~\ref{orbit} and perform a period search on them. No clear and
outstanding peak is found in the range 20~days~$\le P
\le$~1000~days. However, several weak periods are visible. The
strongest are 6667 (0.000150), 2950 (0.000339), 1626 (0.000615),
301 (0.00332), 216 (0.00463), 203 (0.00493), 130 (0.00769) and
96\,days (0.0104\,c/d). After pre-whitening, i.e. subtracting the
f$_0$=0.000150\,c/d signal from the residual RVs, the power spectrum
does not show any distinct period any more. When subtracting any
of the other periods, the 6667\,d period remains intact. This
means that the 6667\,days period is the only signal present in the
residual data. The next two peaks can be explained as the second
and third peak 2f$_0$ and 4f$_0$ of the 6667d period, the other ones
come from the structure of the window function. Of course, this
period is very close to the total length of our data sample and
therefore uncertain and possibly spurious.

Assuming a third component as the origin of these variations, we can 
calculate an orbit of this potential tertiary star which would be 
surrounding the inner binary system in a wide orbit with a period of 
around 20 years and an amplitude $K_\mathrm{1-2}$ of $\approx 4 \pm 1$\,\kms, 
leading to a $f(\mbox{M})$ of $0.044 \pm 0.035$\,\Msun\ and a $a\nm{1$-$2} \sin i$ 
of $3.7 \pm 1.0\cdot 10^8$\,km\ (Table~\ref{tab:tertiary}, left column; the
right column presents an alternative eccentric orbit). 
All three values shrink in case of non-zero eccentricity.
In \S\ref{sec:mass}, we will discuss the mass of a potential tertiary.

\section{Astrophysical parameters of EI~Eri}
\subsection{Effective temperature}\label{chap:spectype}\label{chap:teff}
\citet{bidelman73} classified EI\,Eri as G5\,{\sc iv} from
objective-prism plates. \citet{harlan74} classified it as G2\,{\sc
iv-v}, while \citet{abt86} assumed a ``weak-lined'' G5\,{\sc v}
classification from photometric indices. \citet{cutispoto95} finds
its UBVRI colors consistent with a G5\,{\sc iv} + G0\,{\sc v}
system. The classification of G5\,{\sc iv} is also favored by
\citet{fekel:moffett86} on the basis of high-resolution spectra.

The effective temperature of EI\,Eri can be estimated by comparing the observed
$U - B$, $B - V$, $V - R_C$ and $V - I_C$ colours \citep[which are 0\fm13, 0\fm67,
0\fm40 and 0\fm77, respectively;][]{cutispoto95} to synthetic and empirical values
from colour-temperature transformation lists.
For EI\,Eri's $B - V$ value of 0\fm{67},
\citet{flower96} lists a temperature of 5653\,K.
Comparing with ATLAS-9 model atmospheres
\citep{kurucz93}\footnote{http://ccp7.dur.ac.uk/ccp7/Atlas/colours/ubvm05k2.dat},
we find our $U - B$ and $B - V$ values to be consistent with a temperature of \Teff = 5500\,K
for metallicities between -0.5 and -1.0~dex
which is required by the reconstruction of the line profiles (see paper {\sc ii}).
Solar abundances would yield 250\,K higher temperature.
$V - I_C$ and $V - R_C$ yield temperatures in-between 5500 and 5750\,K.
\citet{schmidt-kaler} lists 5770\,K for $B - V$ = 0\fm{68} for luminosity class {\sc v}.
\citet{randich:gratton93} derived a \Teff\ of 5700\,K using a spectrum synthesis analysis.
\citet{nordstrom:mayor04} list an effective temperature of 5408\,K 
\citep[from the calibration of][]{alonso:arribas96}.
We finally adopt a value of 5500\,K which is supported by our Doppler imaging results
\citep{kgs:rice91,wasi01,paper2}.

The above temperatures are formally for the combined EI\,Eri binary 
star. However, no spectral lines from the secondary are seen which means 
that the continuum ratio must be larger than about a factor 10. 
According to our orbit, $M_2$ amounts to $0.23 \pm 0.04$\,\Msun\ (see 
\S\ref{sec:mass}) which corresponds to a spectral type of \type{M4-5}{v} 
and to $L_2 \approx 1$--$2\cdot 10^{-2}$\,\Lsun\ (assuming a dwarf). 
Compared to the primary ($L_1 = 4.6$\,\Lsun), the secondary contributes 
only 2--4\permil\ of the total light and can therefore be neglected for the Doppler imaging study.

\begin{table}[t]
\caption{Original and improved astrophysical parameters of EI\,Eri. Orbital values are
listed separately in Table~\ref{tab:orbit}.}\label{tab:param:improved}
\begin{center}
\begin{tabular}{@{}l@{\,}l@{\,}l@{}}
\hline
\noalign{\smallskip}
Parameter & Original & Improved value \\
          &  value   & (this paper)   \\
\noalign{\smallskip}
\hline
\noalign{\smallskip}
Spectral type         & \type{G5}{iv}      $^a$   & \type{G5}{iv} + (\type{M4-5}{v}) $^b$ \\
\Teff            $[$K$]$  & 5460 $^c$, 5600 $^d$  & 5500 $\pm$ 100    \\
$T_\mathrm{spot}$ $[$K$]$ & $3600\pm400$   $^c$   & 3600 $^m$         \\
                          & $3700\pm150$   $^d$   & \dots             \\
$v \sin i$     $[$\kms$]$ & 50             $^e$   & $51 \pm 0.5$      \\
\Porb          $[$days$]$ & 1.947227\,(8)  $^c$   & $1.947232 \pm 0.000004$ \\
$P\nm{phot}$   $[$days$]$ & 1.945          $^f$   & $1.95272 \pm 0.00003$ \\
                          & 1.9527         $^g$   & \dots             \\
                          & 1.952717\,(31) $^h$   & \dots             \\
\MV            $[$mag$]$  & 2.75           $^i$   & $3.20 \pm 0.12$   \\
$L_1$, $L_2$   $[$L$_{\odot}$$]$ & --             & $4.60 \pm 0.35$, (0.01-0.02) $^b$ \\
$R_1$, $R_2$   $[$\Rsun$]$& $\ge$\,1.9     $^i$   & $2.37 \pm 0.12$, (0.3) $^b$ \\
$M_1$, $M_2$   $[$\Msun$]$& $\ge 1.4$, 0.53$-$1.3 $^i$~ & $1.09\pm 0.05$, $0.23\pm 0.04$ \\
Inclination    $[$\degr$]$&$34 \le i \le 58$ $^i$ & $56.0 \pm 4.5$    \\
Parallax       $[$mas$]$  & $17.80 \pm 0.97$ $^j$ & --                \\
Distance       $[$pc$]$   & $56.2 \pm 3.1$   $^j$ & --                \\
Ang.\,diam.\,$[$mas$]$    & 0.4--0.6         $^k$ & $0.3921 \pm 0.0006$ \\
Age            $[$Gyr$]$  & --                    & $6.15 \pm 0.50$     \\
\noalign{\smallskip}
\hline
\noalign{\smallskip}
\multicolumn{3}{l}{$^a$ \dots\ \citet{cutispoto95}       }\\
\multicolumn{3}{l}{$^b$ \dots\ according to $M_2 = 0.23$\,\Msun\ from \S\ref{sec:mass} and the spec-}\\
\multicolumn{3}{l}{~~~~~~~~\,tral type-luminosity tables in \citet{schmidt-kaler}.}\\
\multicolumn{3}{l}{$^c$ \dots\ \citet{kgs90}             }\\
\multicolumn{3}{l}{$^d$ \dots\ \citet{oneal:neff98}      }\\
\multicolumn{3}{l}{$^e$ \dots\ \citet{fekel:moffett86}   }\\
\multicolumn{3}{l}{$^f$ \dots\ \citet{hall:osborn87}     }\\
\multicolumn{3}{l}{$^g$ \dots\ Olah et al. (2000, 2002)  }\\
\multicolumn{3}{l}{$^h$ \dots\ \citet{kgs:bartus97}      }\\
\multicolumn{3}{l}{$^i$ \dots\ \citet{fekel:quigley87}   }\\
\multicolumn{3}{l}{$^j$ \dots\ \citet{hip}               }\\
\multicolumn{3}{l}{$^k$ \dots\ \citet{wittkowski:schoeller02}}\\
\multicolumn{3}{l}{$^m$ \dots\ see paper~{\sc ii} of this series}\\
\end{tabular}
\end{center}
\end{table}

\subsection{Rotational broadening and stellar radius}\label{chap:vsini}\label{sec:v+Rsini}
The projected equatorial velocity, $v \sin i$, was estimated by
\citet{fekel:moffett86} to be $50 \pm 3$ \kms\ while
\citet{hatzes:vogt92} and \citet{donati:semel97} gave $50 \pm 1$
\kms\ and $51 \pm 1$~\kms, respectively. For the present
investigation, we use the Doppler-imaging technique to verify and
slightly improve the \vsini value. Doppler imaging is very
sensitive to the projected rotational velocity and provides better
accuracy than any other method, as it takes into account the line
deformation due to spots. Tests with images of the $\lambda$6393,
$\lambda$6411 and $\lambda$6439 line using the Doppler imaging
codes {\sc TempMap} and {\sc DOTS} lead to a value of $51.0 \pm
0.5$~\kms. See paper~{\sc ii} of this series.

The minimum stellar radius
$R \sin i = (P\nm{rot} \cdot v \sin i)/50.6$
with $P\nm{rot} = P\nm{orb} = 1.9472324 \pm 0.0000038$
is $1.963 \pm 0.019$~\Rsun.\footnote
{~Compare: For $P\nm{rot} = P\nm{phot} = 1.9527 \pm 0.0001$, we get
        $R \sin i = 1.968 \pm 0.019\ \Rsun$.
	In paper {\sc iii}, we will address the ditferencent values of
	the photometric and orbital period.}
Obviously, EI\,Eri's primary star is a subgiant rather than a dwarf,
ruling out the luminosity class {\sc v} suggested earlier by \citet{abt86}.

\begin{figure*}
\begin{center}
\includegraphics[width=81mm,angle=0,clip]{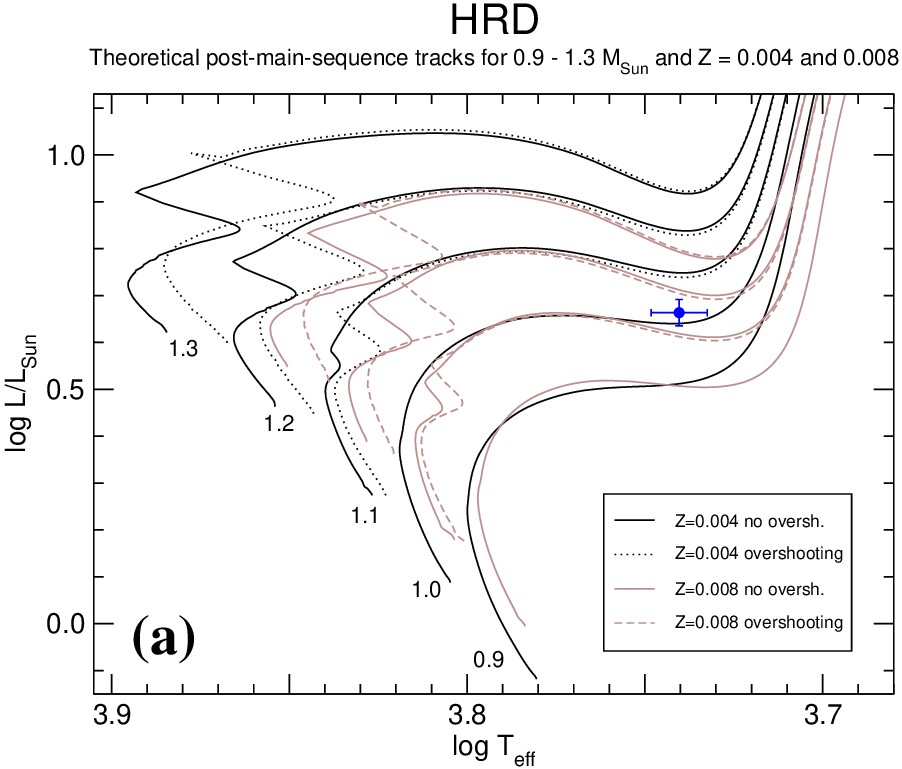}
\includegraphics[width=84mm,angle=0,clip]{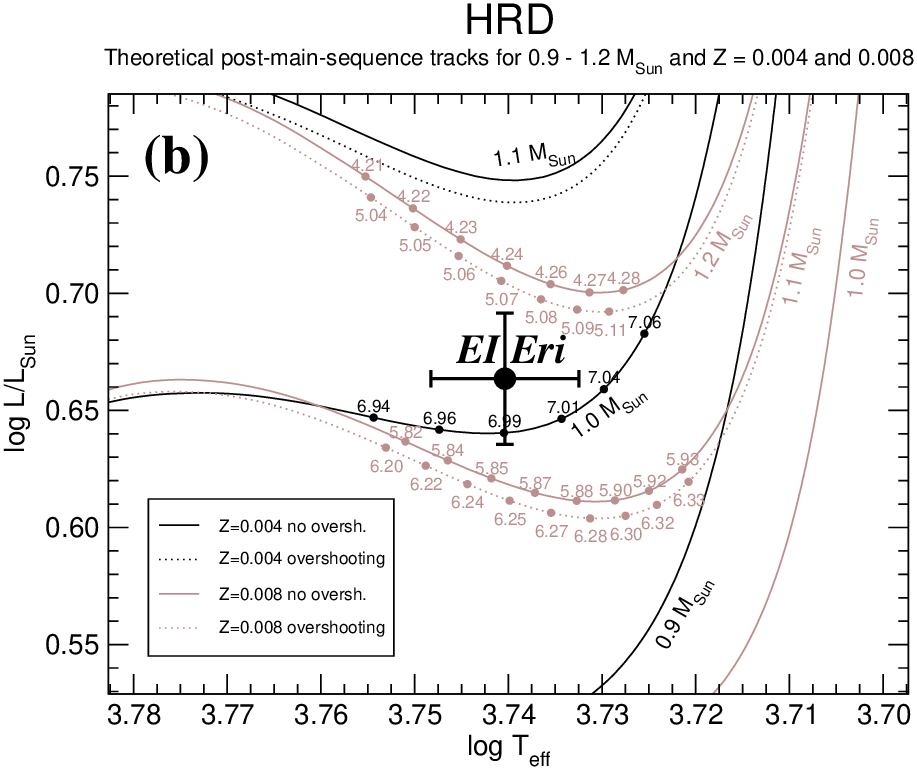}
\caption{{\bf (a)} Evolutionary tracks for post-main-sequence stars from
\citet{pietrinferni:cassisi04} for masses of 0.9 -- 1.3~\Msun\ and for two different metallicities.
The position of EI\,Eridani is marked at $\log \Teff = 3.740 \pm 0.008$ ($5500 \pm 100$\,K)
and $\log L/\Lsun = 0.66 \pm 0.03$ ($4.6 \pm 0.3$\,\Lsun). {\bf (b)} shows the central
range of (a) but adding age along the evolutionary tracks (the values are Gyr).
For a metallicity of Z = 0.006, these
tracks suggest a primary mass of $M_1 = 1.09 \pm 0.05$\,\Msun\ and an age
of $6.15 \pm 0.50$\,Gyr.}\label{evolstat}
\end{center}
\end{figure*}

\subsection{Luminosity and radius}\label{sec:hip}
The {\em Hipparcos} spacecraft measured a trigonometric parallax
of 17.80$\pm$0.97 milli-arcsec \citep{hip}, corresponding to a
distance of 56.2$\pm$3.1~pc. The brightest (least-spotted) $V$
magnitude of EI~Eri was observed on December 18, 2001, by the
Vienna APTs and corresponds to 6\fm949. We adopt this as the least
spotted $V$ brightness.
Combined with the distance, this gives \MV\ = 3\fm20$\pm$0.12.
According to \citet{flower96}, the bolometric correction for a
subgiant with a temperature of $5500 \pm 100$~K is $-0.139 \pm
0.025$. Therefore, the bolometric magnitude is $3\fm061 \pm 0.082$
(neglecting interstellar absorption as it is a nearby star). As we
cannot adopt an external error for the maximum brightness the
above error merely reflects the error of the parallax. However, if
the brightest visual magnitude is smaller than 6\fm9, it does not
affect the error of the stellar radius, but just the radius
itself.

Adopting a solar bolometric magnitude of 4\fm72, we get the luminosity $L/L_\mathrm{\odot} =
4.60 \pm 0.35$. Using $\Teff = 5500 \pm 100$\,K, we obtain a stellar radius of
$2.37 \pm 0.12$\,R$_{\odot}$ from the Stefan-Boltzmann law (independent of the inclination)
and consistent with the value of $R/R_{\odot} \geq 1.9$ from \citet{fekel:quigley87}.
This radius corresponds to an angular diameter of $\theta = 0.3921 \pm 0.0006$\,mas,
confirming the estimations from \citet{wittkowski:schoeller02}. 

\subsection{Orbital inclination}\label{sec:inclination}
The analysis by \citet{fekel:quigley87} suggested an inclination
between $15\degr \le i \le 58\degr$.\footnote{Note that
\citet{fekel:quigley87} accidently used $f(\mbox{M}) = 0.041$
instead of 0.0041 as listed in their paper. Accordingly, the
values in their Table~{\sc V} are wrong. The correct inclination
limits using their data values should read $15\degr \le i \le
58\degr$ instead of $34\degr \le i \le 58\degr$.} We derive a
primary mass of $M_1 = 1.09\pm 0.05$\,\Msun\ from \Teff\ and the parallax
(see \S\ref{sec:mass}; both values are independent of the
inclination). Using the observed mass function $f(\mbox{M}) =
0.00391 \pm 0.00010$ (see Table~\ref{tab:param:improved}), we
achieve a minimum mass for the secondary (assuming $i = 90\degr$)
of $M_2 \ge 0.184 \pm 0.008$\,\Msun. This gives a maximum mass
ratio of $M_1$/$M_2 \le 5.91 \pm 0.25$ and an $a \sin i \le 4.97
\pm 0.25 \cdot 10^6$\,km. Because EI\,Eri shows no eclipses (i.e.,
$R_1 + R_2 < a\cos i$), we can use our maximum value for $a \sin i$ and the
radius $R_1$ and achieve maximum values for the orbital inclination:
$i \le 70\fdg9 \pm 2\fdg0$ for $R_2 = 0.1$~\Rsun, $i \le 69\fdg5
\pm 1\fdg9$ for $R_2 = 0.3$~\Rsun, $i \le 68\fdg1 \pm 1\fdg9$ for
$R_2 = 0.5$~\Rsun\ or $i \le 66\fdg7 \pm 1\fdg9$ for $R_2 =
0.7$~\Rsun. Using $\sin i = (\Prot \cdot v \sin i) / (50.6 \cdot
R/\Rsun)$ and taking \Prot = \Porb, we obtain $i = 56.0 \pm 4.5
\degr$ in agreement with any of the above secondary radii.
Assuming the orbital and the stellar rotational axis to be
perpendicular, we adopt this as our final inclination value.

\subsection{Mass and evolutionary status}\label{sec:mass}\label{sec:evolstat}
\citet{fekel:quigley87} estimated $1.4 \le M_1/\Msun\le 1.8$ for the primary and
an upper limit for the secondary (for $i = 90\degr$) of 1.0 -- 1.3 $M_1/\Msun$.
They suggested the secondary to be a late K or early M dwarf (since no evidence for a hot white
dwarf companion is seen in the ultraviolet).

With the relatively precise luminosity from \S\ref{sec:hip} and
the effective temperature from \S\ref{chap:spectype}, we can
compare the position of EI\,Eri's primary in the  H-R diagram with
theoretical evolutionary tracks. We use the scaled solar models from
\citet{pietrinferni:cassisi04}.
As argued in paper {\sc ii}, we assume abundances to be around $-0.50$\,dex 
below solar, corresponding to a metallicity value of Z = 0.006.
The closest models by \citet{pietrinferni:cassisi04} have Z = 0.004 and 
0.008 and are displayed in Fig.~\ref{evolstat}.
The former corresponds to a mass of 1.02\,\Msun\ and an age of 7.0~Gyr, 
the latter requires interpolation between the 1.1 and 1.2 solar-mass cases 
and results in 1.15\,\Msun\ and 5.0~Gyr (with overshooting) and 
1.16\,\Msun\ and 5.6~Gyr (without overshooting).\footnote{For Z = 0.004 
and M = 1.0\,\Msun, only canonical models without overshooting are 
available.} Interpolating both cases results in values of $1.09 \pm 
0.05$\,\Msun\ and $6.15 \pm 0.50$\,Gyr.
Comparing these results with the findings of \citet{nordstrom:mayor04}, 
we see that mass is in excellent agreement with their value of 
$1.10^{+0.10}_{-0.12}$\,\Msun\ while age is well within the error bars 
of $5.5^{+1.2}_{-1.1}$~Gyr. \citet{demircan:eker06}, using old orbital data,
give a total mass of 1.18\,\Msun\ and an age of 9.16\,Gyr.

With the mass function $f(\mbox{M}) = (M_2 \sin i)^3 / (M_1 
+ M_2)^2 = 0.00391$ (see Table~\ref{tab:orbit}) and $i = 56\fdg0 
\pm 4\fdg5$ (see \S\ref{sec:inclination}), the primary mass of 
$1.09 \pm 0.05$\,\Msun\ requires the mass of the secondary to be $0.228 
\pm 0.044$\,\Msun\ corresponding to $K_2 = 128 \pm 25$\,\kms\ and $a_2 
\sin i = 3.4 \pm 0.7 \cdot 10^6$\,km.

From the combined mass for the binary system ($M_1 + M_2$) of $1.32 \pm 
0.07$\,\Msun, we can incorporate the mass function for the proposed 
binary/tertiary system (see \S\ref{sec:triple}) and repeat the above 
evaluation in order to estimate the tertiary mass $M_3$. A period of 
$6700 \pm 500$ days yields (for the circular orbit): $M_3$ = $0.7 \pm 
0.6$\,\Msun\ corresponding to $K_3 = 8 \pm 7$\,\kms\ and $a_3 \sin i = 7 
\pm 7 \cdot 10^8$\,km (again for $i = 56\fdg0 \pm 4\fdg5$) with $a$ 
being $9 \pm 6$~AU (corresponding to an angular separation from the binary
of 0.3 arc sec). The large errors are mainly due to the large error in the mass 
function of the binary-tertiary orbit which itself is mainly due to the 
large error in $K_{1-2} = 4 \pm 1$\,\kms, the half-amplitude of the 
barycenter variation.
The appropriate values for the eccentric orbit are $M_3$ = $1.1 \pm 1.7$\,\Msun,
$K_3 = 7 \pm 11$\,\kms, $a_3 \sin i = 6 \pm 10 \cdot 10^8$\,km and $a = 9 \pm 8$~AU.

\subsection{Roche lobe and gravity}\label{sec:roche}\label{sec:logg}
\begin{figure}
\centering
\includegraphics[width=83mm,clip,angle=0]{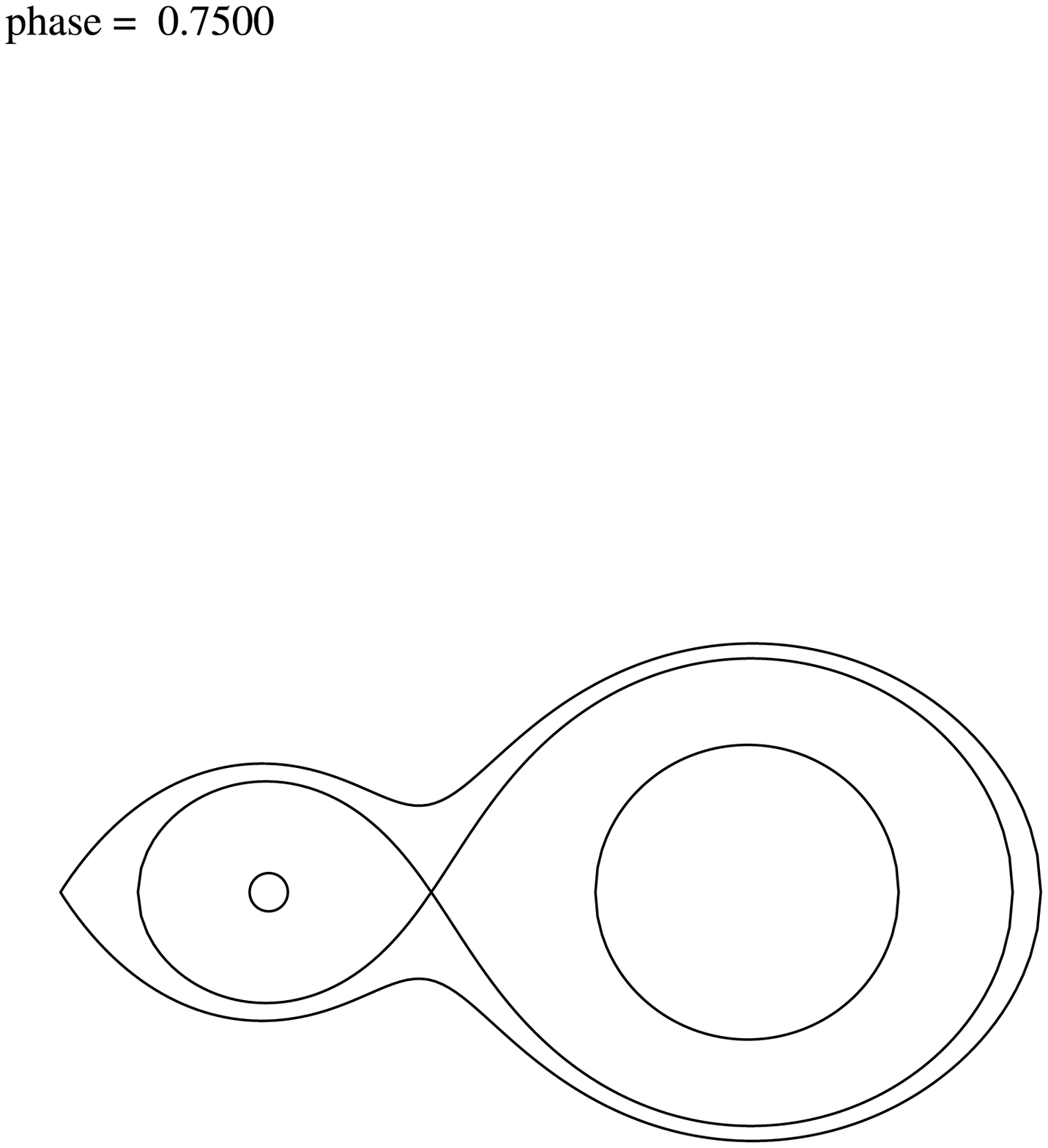}
\caption{Critical Roche equipotentials of the EI\,Eri system. The
inner circles are the stellar surfaces, the other lines denote the
inner and outer critical equipotentials, respectively. The
\type{G5}{IV} primary fills $\approx 64\%$ of its Roche lobe and
is practically spherical in shape.} \label{rochelobe}
\end{figure}

The Roche-lobe radius is determined from the semi-major orbital axis and the mass ratio.
Assuming $M_1$/$M_2 = 4.77 \pm 0.05$\,\Msun\ and
$a \sin i = a_1 \sin i + a_2 \sin i = 4.145\pm 0.725\cdot 10^6$\,km,
we can, using the formula from \citet{eggleton83}, calculate the Roche-lobe radii
$R\nm{L,1} \sin i = 3.08 \pm 0.52$\,\Rsun\ and
$R\nm{L,2} \sin i = 1.52 \pm 0.26$\,\Rsun.
Using $i = 56\fdg0 \pm 4\fdg5$, this translates to
$R\nm{L,1} = 3.71 \pm 0.62$\,\Rsun\ and
$R\nm{L,2} = 1.83 \pm 0.32$\,\Rsun\
for the primary and the secondary, respectively. This means that the 
primary fills $\approx 64 \pm 12\%$ of its Roche lobe but is still 
significantly detached from its inner critical equipotential surface. 
The secondary with assumed $R_2 \approx 0.3$\,\Rsun\ 
\citep[for a \type{M4-5}{v};][]{schmidt-kaler} fills just 20\% of 
its Roche lobe.
Figure~\ref{rochelobe} shows a graphical view of the binary system. 
Shown are the location of the stellar surfaces and the inner and outer 
critical equipotentials as obtained with the program BinaryMaker 
\citep{bradstreet93}.

\citet{oneal:saar96} list a gravity of $\log g \approx 3.8$ consistent
with $\log g = 3.75$ used by \citet{hackman:piskunov91}.
\citet{randich:gratton93} determined a $\log g$ of 4.1 using
spectrum synthesis analysis. \citet{gray} lists a $\log g$ of 4.46
for a G5 dwarf (V) and 3.3 for a G5 giant (III).
Using the mass and radius of the primary from the previous analysis and applying the relation
$g = G M R^{-2}$, where $G$ is the gravitational constant, we can directly calculate
$\log g$ to $3.73 \pm 0.07$, in good agreement with previous claims.

\subsection{Lithium abundance}\label{sec:lithium}
\begin{figure}
\centering
\includegraphics[width=83mm,clip,angle=0]{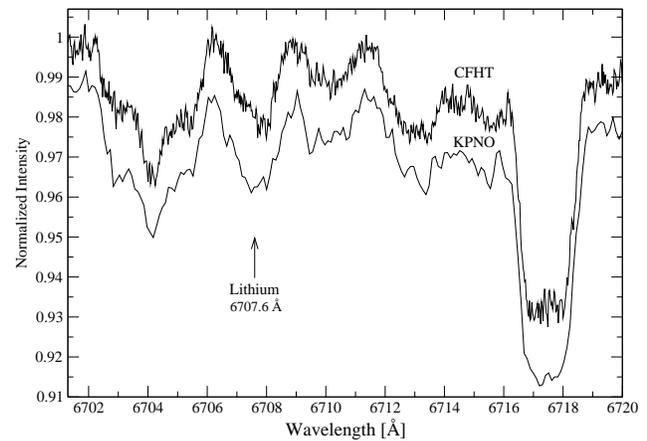}
\caption{Lithium spectra from CFHT and KPNO. The KPNO spectrum is shifted by $-$0.01 in
intensity for better viewing. Signal-to-noise is $\approx$~550/330 (CFHT/KPNO),
dispersion 0.027/0.103\,\AA, integration time 10/60\,min.}
\label{lithium}
\end{figure}

Measurements of two spectra in the lithium region (one from CF96 and
one from a CFHT observing run in October 1997; see
Fig.~\ref{lithium}) give an equivalent width of $40 \pm 5$\,m\AA\ in
good agreement with the value obtained by \citet{fekel:quigley87} of
36\,m\AA. Comparing this to theoretical curves of growth from
\citet{pavlenko:magazzu96}, both LTE and non-LTE, we derive a lithium
abundance of $\log n$(Li) = 2.0 for \Teff = 5500\,K and $\log g$ = 3.5--4.5.
If the equivalent width is reduced to 25\,m\AA\ to
account for the Fe and V blends, the abundance is still $\log n$(Li) =
1.75. \citet{randich:gratton93} estimate $\log n$(Li) = 1.8 using
spectrum synthesis analysis.

\section{Summary}
The following results were obtained:
\begin{itemize}
\item
Improved orbital parameters were determined with HJD =
2,448,054.7109 + 1.9472324 $\times\ E$, $\gamma$ = 21.6\,\kms, $K_1$
= 26.8\,\kms. The barycenter velocity $\gamma$ exhibits variations
of about 8\,\kms\ that are possibly cyclic. We interpret these
variations due to a third stellar component and suggest a possible
orbital solutions for the tertiary with a period of around 6700 days
(19 years) and an amplitude of 4 -- 6\,\kms. An instrumental origin
cannot be definitely ruled out. However, a thorough investigation of
the orbit solutions of the individual observing runs and their
radial velocity standard stars confirmed its reality in our data.
The proposed period is also supported by the smaller {\em O--C}
radial velocity residuals. The third component would thereby
surround the known binary system in a wide orbit. Corresponding
values for the tertiary mass and the orbital parameters of the
binary--tertiary system are quoted (Tab~\ref{tab:tertiary}).
However, its plausibility cannot be asserted unless a recurrence
is observed or higher-quality RV measurements are available.
\item
From the \emph{Hipparcos} parallax, we obtain a luminosity of $4.60
\pm 0.35$\,\Lsun\ and a radius of $2.37 \pm 0.12$\,\Rsun, independent
from the inclination.
\item
With our new parameters and by using evolutionary tracks from 
\citet{pietrinferni:cassisi04}, we locate the position of the primary in 
the HR diagram and determine its mass to $1.09 \pm 0.05$\,\Msun\ and an 
age of $6.15 \pm 0.50$\,Gyr.
\item
The inclination is found to be $i = 56.0 \pm 4.5\degr$.
\item
Using the observed mass function of the inner binary and assuming
$i\nm{orb} \equiv i\nm{*}$, we derive the mass of the unseen secondary:
$M_2 = 0.23 \pm 0.04$\,\Msun\ corresponding to an absolute
magnitude of \MV $\approx 12 \pm 1$ \citep{baraffe:chabrier98}.
\end{itemize}

\acknowledgements We are very grateful to the Deutsche
Forschungsgemeinschaft for grant STR 645/1, for the
Hungarian-German Intergovernmental Grant D21/01, and for the Hungarian
Research Grant OTKA T-48961. Special thanks go
to J\'anos Bartus for his support and computer assistance, to
Frank Fekel for valuable discussions about the orbital solution,
and to Johanna Jurcsik for help with the orbital period. This
research project made extended use of the SIMBAD database,
operated at CDS, Strasbourg, France. The orbit program by Barker
and the utility {\tt BinaryMaker} by David Bradstreet were used
for this paper.



\bibliography{aa_mnem,kgs_etal,wasi}

\appendix

\section{Observing logs (Online material)}\label{tab:obslog}
Observing logs for all data used in this investigation, the ones
from our own observing runs as well as the ones from the
literature, are specified in the following tables. The tables list
the Heliocentric Julian Date (HJD) at the midpoint of an
observation, the exposure
time (combined exposure time in case of combined images; actually,
CCD dark time is used), the signal-to-noise ratio as estimated
using {\tt IRAF\/}'s  {\tt splot\/}, the phase, the
heliocentric radial velocity $v_r$ in \kms, and the error of the
radial velocity measurement $\sigma_{v_r}$ in \kms\ (``DI'' means
that $v_r$ was estimated by means of Doppler imaging, and no $\sigma_{v_r}$
can be given).
For the MUSICOS~98 data, the observing site is listed as well, and the
HJD had to be abbreviated (24511+). For the data from the literature,
HJD (244+), phase, $v_r$, the used instrument, and a reference code is given.

\begin{center}

\mbox{NSO, Nov 16 -- Dec 22, 1988}
\begin{supertabular}{ccccrc}
\hline
\noalign{\smallskip}
 HJD & $t\nm{exp}$ & S/N & phase & \multicolumn{1}{c}{$v_r$} & $\sigma_{v_r}$ \\
 & $[$s$]$ & & & \multicolumn{2}{c}{$[$\kms$]$} \\
\noalign{\smallskip}
\hline
\noalign{\smallskip}
2447481.85877934 & 3600 & 280 & 0.8121 &  22.92 & (DI) \\
2447482.72595936 & 2400 & 240 & 0.2575 &  11.59 & (DI) \\
2447485.78022403 & 3000 & 270 & 0.8260 &  24.96 & (DI) \\
2447485.91420547 & 4500 & 270 & 0.8948 &  35.20 & (DI) \\
2447486.78984068 & 2400 & 270 & 0.3445 &  -1.94 & (DI) \\
2447486.92332404 & 2400 & 290 & 0.4130 & -10.62 & (DI) \\
2447492.77544820 & 2400 & 280 & 0.4184 & -10.40 & (DI) \\
2447493.80873462 & 2400 & 270 & 0.9490 &  38.34 & (DI) \\
2447494.86839555 & 1500 & 250 & 0.4932 & -14.11 & (DI) \\
2447508.90138814 & 1800 & 290 & 0.6999 &   3.99 & (DI) \\
2447509.78106126 & 2100 & 270 & 0.1516 &  30.67 & (DI) \\
2447509.80592715 & 1800 & 280 & 0.1644 &  29.19 & (DI) \\
2447511.78858361 & 2400 & 280 & 0.1826 &  24.80 & (DI) \\
2447511.93134400 & 2509 & 200 & 0.2559 &  13.26 & (DI) \\
2447513.77129171 & 1800 & 230 & 0.2008 &  20.90 & (DI) \\
2447513.92290565 & 2100 & 200 & 0.2787 &   8.24 & (DI) \\
2447517.92410247 & 3900 & 250 & 0.3335 &  -0.36 & (DI) \\
\noalign{\smallskip}
\hline
\end{supertabular}
\vspace{3mm}


\mbox{NSO, Jan 12 -- Jan 31, 1989}
\begin{supertabular}{ccccrc}
\hline
\noalign{\smallskip}
 HJD & $t\nm{exp}$ & S/N & phase & \multicolumn{1}{c}{$v_r$} & $\sigma_{v_r}$ \\
\noalign{\smallskip}
\hline
\noalign{\smallskip}
 2447538.81751973 & 3600 & 370 & 0.0633 & 46.11 & (DI) \\
 2447545.69961784 & 1815 & 340 & 0.5976 & -6.03 & (DI) \\
 2447546.67373727 & 1680 & 330 & 0.0978 & 42.21 & (DI) \\
 2447549.68456430 & 1800 & 270 & 0.6440 & -1.07 & (DI) \\
 2447556.69637278 & 2700 & 340 & 0.2449 & 19.06 & (DI) \\
 2447557.69618472 & 2400 & 380 & 0.7584 & 17.01 & (DI) \\
\noalign{\smallskip}
\hline
\end{supertabular}
\vspace{3mm}


\mbox{NSO, Nov 9, 1989, -- Jan 24, 1990}
\begin{supertabular}{ccccrc}
\hline
\noalign{\smallskip}
 HJD & $t\nm{exp}$ & S/N & phase & \multicolumn{1}{c}{$v_r$} & $\sigma_{v_r}$ \\
\noalign{\smallskip}
\hline
\noalign{\smallskip}
2447839.90839991 & 2700 & 310 & 0.6883 & -- & -- \\
2447840.69110908 & 2400 & 260 & 0.0903 & -- & -- \\
2447841.92716837 & 2400 & 280 & 0.7250 & -- & -- \\
2447864.96392067 & 2100 & 240 & 0.5556 & -- & -- \\
2447865.94967414 & 2400 & 320 & 0.0618 & -- & -- \\
2447878.94827398 & 3000 & 150 & 0.7372 & -- & -- \\
2447898.78161089 & 2400 & 200 & 0.9226 & -- & -- \\
2447899.77544479 & 2100 & 200 & 0.4330 & -- & -- \\
2447900.77487947 & 1800 & 310 & 0.9463 & -- & -- \\
2447900.80027108 & 2100 & 280 & 0.9593 & -- & -- \\
2447901.64796707 & 1200 & 280 & 0.3946 & -- & -- \\
2447901.66231884 & 1200 & 240 & 0.4020 & -- & -- \\
2447902.73509105 & 1200 & 240 & 0.9529 & -- & -- \\
2447902.74944197 & 1200 & 250 & 0.9603 & -- & -- \\
2447903.72759042 & 1200 & 230 & 0.4626 & -- & -- \\
2447903.74192943 & 1200 & 240 & 0.4700 & -- & -- \\
2447912.81237266 & 1800 & 210 & 0.1281 & -- & -- \\
2447915.68691188 & 1800 & 240 & 0.6043 & -- & -- \\
2447925.76718365 & 2400 & 200 & 0.7810 & -- & -- \\
2447925.81426603 & 2700 & 230 & 0.8052 & -- & -- \\
\noalign{\smallskip}
\hline
\end{supertabular}
\vspace{3mm}


\mbox{NSO, Oct 23 -- Dec 2, 1990}
\begin{supertabular}{ccccrc}
\hline
\noalign{\smallskip}
 HJD & $t\nm{exp}$ & S/N & phase & \multicolumn{1}{c}{$v_r$} & $\sigma_{v_r}$ \\
\noalign{\smallskip}
\hline
\noalign{\smallskip}
2448187.90907048 & 2100 & 280 & 0.4038 &  1.51 & 2.17 \\
2448188.88383156 & 1800 & 350 & 0.9044 & 44.40 & 2.31 \\
2448189.80971259 & 1800 & 280 & 0.3799 &  8.86 & 5.90 \\
2448199.04278572 & 2280 & 320 & 0.1216 &  --   &  --  \\
2448200.03172236 & 3600 & 240 & 0.6294 &  4.80 & 1.73 \\
2448204.01902843 & 4200 & 370 & 0.6771 & 10.62 & 1.86 \\
2448211.90627687 & 1800 & 330 & 0.7276 & 18.67 & 1.71 \\
2448214.92815164 & 2400 & 310 & 0.2795 & 17.85 & 2.05 \\
2448224.00487253 & 3300 & 300 & 0.9408 & 49.40 & 2.16 \\
2448227.84227865 & 3000 & 340 & 0.9115 & 45.78 & 1.58 \\
\noalign{\smallskip}
\hline
\end{supertabular}
\vspace{3mm}


\mbox{NSO, Jan 15 -- Jan 31 / Mar 8, 1991}
\begin{supertabular}{ccccrc}
\hline
\noalign{\smallskip}
 HJD & $t\nm{exp}$ & S/N & phase & \multicolumn{1}{c}{$v_r$} & $\sigma_{v_r}$ \\
\noalign{\smallskip}
\hline
\noalign{\smallskip}
2448271.80021084 & 2700 & 300 & 0.4861 & -3.45 & 1.82 \\
2448272.87407843 & 3000 & 230 & 0.0376 & 48.61 & 3.50 \\
2448274.79259064 & 2700 & 360 & 0.0228 & 49.77 & 1.64 \\
2448284.81303141 & 2700 & 390 & 0.1688 & 36.82 & 2.16 \\
2448286.79004258 & 4800 & 390 & 0.1841 & 34.83 & 1.70 \\
2448287.78336189 & 1800 & 360 & 0.6942 & 14.03 & 1.66 \\
2448323.74119835 &  946 & 110 & 0.1603 & 37.02 & 1.98 \\
2448330.66309608 & 3900 & 350 & 0.7151 & 17.80 & 1.98 \\
2448333.67863402 & 4200 & 280 & 0.2637 & 21.20 & 1.61 \\
\noalign{\smallskip}
\hline
\end{supertabular}
\vspace{3mm}


\mbox{NSO, Jan 9 -- Feb 5, 1992}
\begin{supertabular}{cccc@{\hspace{1mm}}rc}
\hline
\noalign{\smallskip}
 HJD & $t\nm{exp}$ & S/N & phase & \multicolumn{1}{c}{$v_r$} & $\sigma_{v_r}$ \\
\noalign{\smallskip}
\hline
\noalign{\smallskip}
2448536.84220564 & 2400 & 300 & 0.5982 &  2.94 & 1.89 \\
2448571.87915138 & 2400 & 300 & 0.5915 &  1.38 & 2.01 \\
2448572.73822549 & 2100 & 250 & 0.0326 & 50.73 & 1.69 \\
2448573.96871468 & 2100 & 300 & 0.6645 & 10.30 & 1.62 \\
2448583.76049664 & 2700 & 250 & 0.6931 & 12.33 & 1.67 \\
2448583.78916548 & 1500 & 300 & 0.7078 & 15.78 & 1.58 \\
2448584.73695026 & 2400 & 350 & 0.1946 &(32.59)&(1.54)\\
2448585.90944273 & 1800 & 300 & 0.7967 & 32.25 & 2.12 \\
2448586.88967786 & 1800 & 300 & 0.3001 & 15.67 & 1.84 \\
2448587.69108233 & 1800 & 300 & 0.7117 & 21.44 & 1.98 \\
2448594.77324122 & 3600 & 350 & 0.3487 &  8.40 & 2.08 \\
2448597.73887879 & 2400 & 300 & 0.8717 & 42.13 & 1.96 \\
2448598.95181001 & 3900 & 250 & 0.4946 & -2.56 & 2.02 \\
\noalign{\smallskip}
\hline
\noalign{\smallskip}
2448630.80189166 & 4500 & 300 & 0.8512 & 39.14 & 1.81 \\
2448631.73144606 & 4200 & 375 & 0.3285 &  9.93 & 2.00 \\
2448635.74885926 & 4200 & 300 & 0.3917 &  2.60 & 1.96 \\
2448644.72706482 & 2400 & 375 & 0.0025 & 51.21 & 1.93 \\
2448645.75364952 & 2700 & 350 & 0.5296 & -2.93 & 1.79 \\
2448646.79759322 & 2700 & 325 & 0.0658 & 48.68 & 2.17 \\
2448647.72925336 & 4500 & 350 & 0.5442 & -3.29 & 2.15 \\
2448657.74044939 & 3300 & 325 & 0.6854 & 14.11 & 2.59 \\
\noalign{\smallskip}
\hline
\end{supertabular}
\vspace{3mm}


\mbox{NSO, Oct 14 -- Dec 1, 1992}
\begin{supertabular}{ccccrc}
\hline
\noalign{\smallskip}
 HJD & $t\nm{exp}$ & S/N & phase & \multicolumn{1}{c}{$v_r$} & $\sigma_{v_r}$ \\
\noalign{\smallskip}
\hline
\noalign{\smallskip}
2448909.81574167 & 3300 & 350 & 0.1386 & 38.89 & 1.80 \\
2448910.82189280 & 2700 & 320 & 0.6553 &  9.57 & 2.53 \\
2448911.76449788 & 2100 & 220 & 0.1393 & 39.80 & 2.90 \\
2448919.01267057 & 2520 & 280 & 0.8616 & 41.80 & 2.41 \\
2448919.80544506 & 2400 & 250 & 0.2688 & 19.80 & 1.84 \\
2448921.92832595 & 1920 & 320 & 0.3590 &  5.41 & 1.47 \\
2448932.98758668 & 3000 & 300 & 0.0384 & 49.89 & 6.25 \\
2448933.85932839 & 1800 & 350 & 0.4861 & -3.52 & 1.40 \\
2448934.97814509 & 2400 & 280 & 0.0607 & 49.12 & 1.87 \\
2448935.97845108 & 4200 & 360 & 0.5744 &  2.31 & 1.53 \\
2448947.99180786 & 4500 & 140 & 0.7439 & 24.23 & 1.72 \\
2448954.93973981 & 3300 & 320 & 0.3120 & 10.72 & 1.07 \\
2448955.91696006 & 2961 & 130 & 0.8138 & 35.42 & 1.97 \\
2448957.74816912 & 3300 & 380 & 0.7542 & 26.90 & 1.76 \\
\noalign{\smallskip}
\hline
\end{supertabular}
\vspace{3mm}


\mbox{NSO, January 29 -- April 2, 1993}
\begin{supertabular}{ccccrc}
\hline
\noalign{\smallskip}
 HJD & $t\nm{exp}$ & S/N & phase & \multicolumn{1}{c}{$v_r$} & $\sigma_{v_r}$ \\
\noalign{\smallskip}
\hline
\noalign{\smallskip}
2447866.97849159 & 2400 &     & 0.5901 & -- & -- \\
2447875.77605125 & 1800 &     & 0.1081 & -- & -- \\
2448958.85377396 & 3000 &     & 0.3220 & -- & -- \\
2449016.74254290 & 4200 & 300 & 0.0508 & 48.96 & 1.53 \\
2449030.68296247 & 2700 & 350 & 0.2098 & 30.22 & 1.63 \\
2449030.70657099 & 1200 & 250 & 0.2220 & 27.71 & 1.93 \\
2449066.66889489 & 4200 & 250 & 0.6904 & 14.14 & 1.54 \\
2449078.66783958 & 1452 & 100 & 0.8525 & 43.72 & 4.22 \\
2449079.66842867 & 2023 & 140 & 0.3663 &  3.87 & 2.57 \\
\noalign{\smallskip}
\hline
\end{supertabular}
\vspace{3mm}


\mbox{NSO, October 20 -- November 4, 1993}
\begin{supertabular}{ccccrc}
\hline
\noalign{\smallskip}
 HJD & $t\nm{exp}$ & S/N & phase & \multicolumn{1}{c}{$v_r$} & $\sigma_{v_r}$ \\
\noalign{\smallskip}
\hline
\noalign{\smallskip}
2449280.92690392 & 2100 & 350 & 0.7225 & 19.75 & 2.35 \\
2449281.99038834 & 2100 & 300 & 0.2686 & 20.99 & 1.70 \\
2449282.81566634 & 3000 & 380 & 0.6924 & 16.17 & 1.60 \\
2449283.03415929 & 1500 & 320 & 0.8046 & 33.03 & 1.57 \\
2449291.85534333 & 1800 & 320 & 0.3348 & 10.40 & 1.15 \\
2449293.00654551 & 1656 & 280 & 0.9260 & 48.62 & 1.53 \\
2449294.82520519 & 1500 & 350 & 0.8599 & 40.48 & 1.41 \\
2449295.85622598 & 1500 & 320 & 0.3894 &  3.78 & 1.20 \\
\noalign{\smallskip}
\hline
\end{supertabular}
\vspace{3mm}


\mbox{NSO, November 11 -- December 9, 1993}
\begin{supertabular}{ccccrc}
\hline
\noalign{\smallskip}
 HJD & $t\nm{exp}$ & S/N & phase & \multicolumn{1}{c}{$v_r$} & $\sigma_{v_r}$ \\
\noalign{\smallskip}
\hline
\noalign{\smallskip}
2449302.86739868 & 1800 & 380 & 0.9900 & 51.32 & 1.83 \\
2449326.88055564 & 1500 & 380 & 0.3219 & 12.57 & 2.19 \\
2449327.67699497 & 1200 & 320 & 0.7309 & 21.65 & 1.49 \\
2449328.74862305 &  900 & 280 & 0.2813 & 18.86 & 1.59 \\
2449329.72085393 & 1500 & 320 & 0.7806 & 29.18 & 1.32 \\
2449330.75974826 & 3000 & 220 & 0.3141 & 10.62 & 1.51 \\
2449330.80034305 & 3000 & 220 & 0.3349 &  9.50 & 1.75 \\
\noalign{\smallskip}
\hline
\end{supertabular}
\vspace{3mm}


\mbox{NSO, January 30 -- February 1, 1994}
\begin{supertabular}{ccccrc}
\hline
\noalign{\smallskip}
 HJD & $t\nm{exp}$ & S/N & phase & \multicolumn{1}{c}{$v_r$} & $\sigma_{v_r}$ \\
\noalign{\smallskip}
\hline
\noalign{\smallskip}
2449382.70823625 & 1800 & 240 & 0.9922 & 51.12 & 1.86 \\
2449382.72958862 & 1800 & 220 & 0.0032 & 51.86 & 1.70 \\
2449383.63808861 & 1200 & 280 & 0.4697 & -2.04 & 1.05 \\
2449383.65247397 & 1200 & 330 & 0.4771 & -1.96 & 1.40 \\
2449384.63884888 & 1800 & 320 & 0.9837 & 50.72 & 1.97 \\
2449384.66018966 & 1800 & 300 & 0.9946 & 50.92 & 1.70 \\
\noalign{\smallskip}
\hline
\end{supertabular}
\vspace{3mm}


\mbox{NSO, November 11 -- December 9, 1993}
\begin{supertabular}{ccccrc}
\hline
\noalign{\smallskip}
 HJD & $t\nm{exp}$ & S/N & phase & \multicolumn{1}{c}{$v_r$} & $\sigma_{v_r}$ \\
\noalign{\smallskip}
\hline
\noalign{\smallskip}
2449416.61073607 & 1500 & 300 & 0.4028 &  4.30 & 1.63 \\
2449417.64027834 & 1800 & 230 & 0.9315 & 49.07 & 1.50 \\
2449418.62555397 & 2700 & 270 & 0.4375 &  0.20 & 2.07 \\
2449427.62140418 & 1800 & 180 & 0.0573 & 49.98 & 1.16 \\
2449427.64274509 & 1800 & 160 & 0.0683 & 49.50 & 1.61 \\
2449427.66438765 & 1800 & 150 & 0.0794 & 48.53 & 1.81 \\
2449427.68571774 & 1800 & 150 & 0.0904 & 47.35 & 3.36 \\
2449428.60637301 & 1800 & 300 & 0.5632 & -0.07 & 1.70 \\
2449429.60576848 & 1800 & 180 & 0.0764 & 47.06 & 1.54 \\
2449429.62775173 & 1800 & 250 & 0.0877 & 46.26 & 1.76 \\
2449429.64908147 & 1800 & 250 & 0.0987 & 45.24 & 1.83 \\
2449429.67041157 & 1800 & 250 & 0.1096 & 43.68 & 2.19 \\
2449429.68829901 & 1200 & 180 & 0.1188 & 42.39 & 2.37 \\
2449430.66364978 & 1630 & 100 & 0.6197 &  4.68 & 2.62 \\
2449439.60910082 & 1500 & 280 & 0.2136 & 31.55 & 1.40 \\
2449440.64563261 & 2700 & 320 & 0.7459 & 23.73 & 1.56 \\
2449440.66718426 &  510 &  60 & 0.7570 & 25.62 & 3.94 \\
2449442.60931421 & 1200 & 250 & 0.7544 & 26.23 & 2.52 \\
2449442.62722388 & 1800 & 320 & 0.7636 & 27.38 & 1.55 \\
2449442.64896526 & 1800 & 280 & 0.7747 & 29.83 & 1.78 \\
2449443.59744365 &  900 & 230 & 0.2618 & 25.33 & 1.62 \\
\noalign{\smallskip}
\hline
\end{supertabular}
\vspace{3mm}


\mbox{NSO, October 18 -- October 31, 1994}
\begin{supertabular}{ccccrc}
\hline
\noalign{\smallskip}
 HJD & $t\nm{exp}$ & S/N & phase & \multicolumn{1}{c}{$v_r$} & $\sigma_{v_r}$ \\
\noalign{\smallskip}
\hline
\noalign{\smallskip}
2449643.94619938 & 2700 & 250 & 0.1508 & 39.03 & 1.70 \\
2449644.73987058 & 2700 & 160 & 0.5584 & -0.98 & 2.37 \\
2449644.77161854 & 2700 & 190 & 0.5747 &  0.09 & 2.58 \\
2449644.97355043 & 2700 & 230 & 0.6784 & 14.64 & 2.57 \\
2449645.71373397 & 2700 &  60 & 0.0585 & 47.97 & 3.02 \\
2449645.74549183 & 2700 & 140 & 0.0748 & 46.68 & 2.45 \\
2449645.77770689 & 2700 & 180 & 0.0914 & 45.39 & 2.06 \\
2449646.90943591 & 2700 & 230 & 0.6726 & 12.77 & 1.76 \\
2449646.94642850 & 3600 & 270 & 0.6916 & 15.48 & 1.88 \\
2449647.94902364 & 3600 & 150 & 0.2065 & 29.23 & 3.38 \\
2449647.99254978 & 3600 & 150 & 0.2288 & 27.75 & 2.40 \\
2449648.03095424 & 2520 & 240 & 0.2485 & 23.85 & 2.31 \\
2449648.98419955 & 2700 & 180 & 0.7381 & 22.43 & 1.40 \\
2449649.01596035 & 2700 & 220 & 0.7544 & 24.85 & 1.66 \\
2449649.04090361 & 1410 & 170 & 0.7672 & 26.65 & 2.47 \\
2449656.86531271 & 3600 & 250 & 0.7854 & 29.86 & 1.84 \\
\noalign{\smallskip}
\hline
\end{supertabular}
\vspace{3mm}


\mbox{NSO, December 17, 1994 -- January 3, 1995}
\begin{supertabular}{ccccrc}
\hline
\noalign{\smallskip}
 HJD & $t\nm{exp}$ & S/N & phase & \multicolumn{1}{c}{$v_r$} & $\sigma_{v_r}$ \\
\noalign{\smallskip}
\hline
\noalign{\smallskip}
2449703.79438644 & 2700 & 250 & 0.8858 & 44.89 & 1.88 \\
2449703.82614415 & 2700 & 230 & 0.9021 & 45.74 & 1.74 \\
2449704.58565054 & 1800 & 220 & 0.2922 & 17.56 & 1.85 \\
2449704.60699218 & 1800 & 210 & 0.3031 & 16.07 & 1.80 \\
2449704.62835689 & 1800 & 180 & 0.3141 & 14.21 & 1.75 \\
2449705.59374487 & 2700 & 280 & 0.8099 & 35.55 & 1.41 \\
2449705.62551423 & 2700 & 280 & 0.8262 & 38.19 & 1.41 \\
2449705.78176925 & 1800 & 300 & 0.9064 & 46.42 & 1.41 \\
2449705.80311061 & 1800 & 320 & 0.9174 & 48.06 & 1.42 \\
2449706.82943194 & 2700 & 340 & 0.4444 & -0.87 & 1.44 \\
2449707.59811282 & 3600 & 310 & 0.8392 & 39.82 & 1.72 \\
2449708.74437943 & 1800 & 330 & 0.4279 &  0.53 & 1.47 \\
2449715.81930526 & 2700 & 280 & 0.0612 & 49.33 & 1.49 \\
2449715.85246988 & 2700 & 170 & 0.0782 & 48.26 & 1.70 \\
2449717.78351050 & 2700 & 270 & 0.0699 & 48.78 & 1.77 \\
2449717.81527916 & 2700 & 320 & 0.0862 & 47.84 & 1.42 \\
2449717.84704816 & 2700 & 300 & 0.1025 & 46.33 & 1.64 \\
2449718.60435442 & 2700 & 150 & 0.4915 & -1.50 & 2.22 \\
2449718.63653990 & 2700 & 220 & 0.5080 & -1.01 & 1.24 \\
2449720.82242596 & 1800 & 200 & 0.6305 &  7.35 & 2.42 \\
2449720.84382622 & 1800 & 200 & 0.6415 &  9.00 & 1.73 \\
\noalign{\smallskip}
\hline
\end{supertabular}
\vspace{3mm}


\mbox{NSO, February 17 -- 20, 1995}
\begin{supertabular}{ccccrc@{\,\,}}
\hline
\noalign{\smallskip}
 HJD & $t\nm{exp}$ & S/N & phase & \multicolumn{1}{c}{$v_r$} & $\sigma_{v_r}$ \\
\noalign{\smallskip}
\hline
\noalign{\smallskip}
2449765.67120109 & 2700 & 100 & 0.6626 & 11.69 &  5.51 \\
2449765.70260143 & 2700 & 110 & 0.6787 & 10.71 & 11.64 \\
2449768.63353196 & 1800 &  80 & 0.1839 & 34.77 &  3.72 \\
2449768.65450472 & 1800 &  70 & 0.1947 & 33.26 &  2.27 \\
2449768.67594000 & 1800 &  70 & 0.2057 & 31.08 &  1.98 \\
\noalign{\smallskip}
\hline
\end{supertabular}
\vspace{3mm}


\mbox{NSO, February 28 -- March 28, 1995}
\begin{supertabular}{ccccrc}
\hline
\noalign{\smallskip}
 HJD & $t\nm{exp}$ & S/N & phase & \multicolumn{1}{c}{$v_r$} & $\sigma_{v_r}$ \\
\noalign{\smallskip}
\hline
\noalign{\smallskip}
2449776.64514955 &  377 &  80 & 0.2983 & 18.69 & 4.60 \\
2449787.61740737 & 2700 &  90 & 0.9331 & 46.40 & 2.24 \\
2449787.64879368 & 2700 & 100 & 0.9492 & 47.24 & 3.63 \\
2449791.64808752 & 2700 & 130 & 0.0030 & 49.53 & 2.82 \\
2449791.67947504 & 2700 & 160 & 0.0191 & 49.56 & 3.18 \\
2449799.64329404 & 3600 & 160 & 0.1089 & 44.23 & 2.67 \\
2449800.63200616 & 2700 &  70 & 0.6167 &  5.99 & 5.68 \\
2449800.66339303 & 2700 &  70 & 0.6328 &  9.19 & 7.10 \\
2449802.63812302 & 2700 & 170 & 0.6469 &  9.02 & 3.56 \\
2449802.66952217 & 2700 & 160 & 0.6631 & 11.74 & 2.89 \\
2449804.62434905 & 2700 & 260 & 0.6670 & 10.16 & 2.89 \\
2449804.65574755 & 2700 & 240 & 0.6831 & 12.08 & 5.45 \\
\noalign{\smallskip}
\hline
\end{supertabular}
\vspace{3mm}


\mbox{NSO, October 1 -- 2, 1995}
\begin{supertabular}{ccccrc}
\hline
\noalign{\smallskip}
 HJD & $t\nm{exp}$ & S/N & phase & \multicolumn{1}{c}{$v_r$} & $\sigma_{v_r}$ \\
\noalign{\smallskip}
\hline
\noalign{\smallskip}
2449992.02679 & 2400 &  50 & 0.9074 & 43.78 & 7.19 \\
2449992.84405 & 3600 &  60 & 0.3271 & 11.40 & 6.06 \\
2449992.89221 & 3600 & 150 & 0.3518 &  8.25 & 4.77 \\
\noalign{\smallskip}
\hline
\end{supertabular}
\vspace{3mm}


\mbox{NSO, November 7 -- December 2, 1995}
\begin{supertabular}{ccccrc}
\hline
\noalign{\smallskip}
 HJD & $t\nm{exp}$ & S/N & phase & \multicolumn{1}{c}{$v_r$} & $\sigma_{v_r}$ \\
\noalign{\smallskip}
\hline
\noalign{\smallskip}
2450029.03403 & 3600 & 220 & 0.9124 & 47.50 & 2.40 \\
2450029.84153 & 3600 & 260 & 0.3271 & 13.59 & 4.46 \\
2450031.97660 & 3600 & 360 & 0.4236 &  0.40 & 1.50 \\
2450032.82948 & 3000 & 380 & 0.8616 & 40.93 & 3.72 \\
2450039.89632 & 3600 & 400 & 0.4907 & -0.87 & 2.41 \\
2450040.82260 & 3600 & 320 & 0.9664 & 53.97 & 3.72 \\
2450040.90672 & 2400 & 330 & 0.0096 & 55.43 & 4.15 \\
2450041.87009 & 2700 & 300 & 0.5044 & -1.25 & 2.50 \\
2450042.80285 & 2700 & 300 & 0.9834 & 49.42 & 3.01 \\
2450042.88547 & 2700 & 320 & 0.0258 & 52.19 & 3.20 \\
2450043.65585 & 2700 & 250 & 0.4214 &  2.36 & 2.23 \\
2450044.85677 & 2100 & 280 & 0.0382 & 51.24 & 4.95 \\
2450051.89799 & 1800 & 320 & 0.6542 &  8.48 & 1.50 \\
2450052.88745 & 1800 & 310 & 0.1623 & 36.68 & 3.20 \\
2450053.80507 & 1800 & 330 & 0.6336 &  6.17 & 1.33 \\
\noalign{\smallskip}
\hline
\end{supertabular}
\vspace{3mm}


\mbox{NSO, December 12 -- 28, 1995}
\begin{supertabular}{ccccrc}
\hline
\noalign{\smallskip}
 HJD & $t\nm{exp}$ & S/N & phase & \multicolumn{1}{c}{$v_r$} & $\sigma_{v_r}$ \\
\noalign{\smallskip}
\hline
\noalign{\smallskip}
2450063.93937 & 3600 & 400 & 0.8380 & 40.14 & 5.86 \\
2450065.87490 & 3600 & 360 & 0.8320 & 36.14 & 2.26 \\
2450066.86716 & 3600 & 340 & 0.3416 &  9.60 & 4.93 \\
2450078.88520 & 3600 & 320 & 0.5135 & -2.43 & 5.57 \\
2450079.89740 & 3600 & 180 & 0.0333 & 52.65 & 2.34 \\
\noalign{\smallskip}
\hline
\end{supertabular}
\vspace{3mm}


\mbox{NSO, Nov 1, 1996 -- Jan 8, 1997}
\begin{supertabular}{ccccrc}
\hline
\noalign{\smallskip}
 HJD & $t\nm{exp}$ & S/N & phase & \multicolumn{1}{c}{$v_r$} & $\sigma_{v_r}$ \\
\noalign{\smallskip}
\hline
\noalign{\smallskip}
2450388.90290475 & 3600 & 150 & 0.7229 & 13.40 & 7.6  \\
2450390.99888385 & 3000 & 130 & 0.7992 & 24.88 & 8.1 \\
2450391.98128049 & 3600 & 150 & 0.3038 & 11.50 & 7.8 \\
2450392.96932388 & 2700 & 130 & 0.8112 & 21.84 & 7.7 \\
2450393.99213488 & 2700 & 160 & 0.3364 & 11.35 & 7.0 \\
2450394.68500721 & 3600 & 150 & 0.6923 & 9.09  & 7.3 \\
2450394.96069560 & 1200 & 160 & 0.8338 & 33.84 & 7.8 \\
2450395.68942407 & 1800 & 140 & 0.2081 & 24.67 & 6.9 \\
2450395.98469570 & 1200 & 160 & 0.3597 & 4.79  & 7.3 \\
2450396.72716891 & 3600 & 160 & 0.7410 & 19.95 & 6.8 \\
2450399.77645195 & 4351 & 140 & 0.3070 & 8.66  & 7.2 \\
2450400.88806165 & 3600 & 130 & 0.8778 & 37.02 & 7.9 \\
2450401.87706386 & 3600 & 140 & 0.3857 & 0.33  & 7.0 \\
2450404.71751004 & 3600 & 150 & 0.8444 & 31.27 & 7.4 \\
2450404.95373214 & 1260 & 120 & 0.9657 & 44.96 & 7.6 \\
2450405.93025135 & 1800 & 140 & 0.4672 & -7.20 & 7.6 \\
2450406.79314622 &  900 & 140 & 0.9104 & 42.55 & 7.0 \\
2450406.92499244 &  900 & 150 & 0.9781 & 44.01 & 7.6 \\
2450408.75429680 & 1260 & 140 & 0.9175 & 43.61 & 8.0 \\
2450408.86176188 &  900 & 120 & 0.9727 & 47.67 & 8.6 \\
2450408.95522238 &  900 & 140 & 0.0207 & 47.40 & 7.7 \\
2450409.75060189 & 1800 & 130 & 0.4292 & -3.87 & 7.3 \\
2450409.87995318 & 1800 & 140 & 0.4956 & -7.74 & 7.4 \\
2450411.84548705 & 2700 & 140 & 0.5050 & -5.61 & 7.2 \\
2450412.86152074 & 2700 & 150 & 0.0268 & 43.92 & 7.2 \\
2450414.75119559 & 2700 & 140 & 0.9972 & 45.59 & 8.4 \\
2450415.78868308 & 2700 & 140 & 0.5300 & -4.54 & 7.4 \\
2450416.84316521 & 2700 & 140 & 0.0716 & 43.77 & 8.0 \\
2450417.87992018 & 3600 & 130 & 0.6040 & -2.33 & 6.7 \\
2450418.86438290 & 1800 & 140 & 0.1096 & 36.45 & 8.9 \\
2450419.70589534 & 1800 & 140 & 0.5417 & -4.57 & 7.6 \\
2450420.68467254 & 1800 & 140 & 0.0444 & 42.15 & 8.0 \\
2450421.80456207 & 1080 & 150 & 0.6195 & -2.76 & 8.2 \\
2450422.79352103 & 1080 & 130 & 0.1274 & 34.40 & 8.5 \\
2450423.78074261 & 1260 & 130 & 0.6343 & 1.02  & 6.8 \\
2450424.90900056 & 1800 & 140 & 0.2138 & 21.38 & 8.4 \\
2450425.88593564 & 1800 & 150 & 0.7155 & 10.92 & 7.2 \\
2450426.89264277 & 1800 & 140 & 0.2325 & 19.08 & 8.6 \\
2450428.91439143 & 1800 & 130 & 0.2707 & 13.82 & 8.2 \\
2450429.87755393 & 1800 & 150 & 0.7654 & 18.65 & 8.0 \\
2450430.88667993 & 1800 & 150 & 0.2836 & 12.93 & 7.8 \\
2450431.71075924 & 1800 & 150 & 0.7068 & 10.53 & 6.9 \\
2450432.89437238 & 1800 & 150 & 0.3146 & 7.32  & 7.7 \\
2450435.73593090 & 2700 & 130 & 0.7739 & 21.00 & 7.0 \\
2450436.72737582 & 2700 & 140 & 0.2831 & 11.22 & 7.6 \\
2450437.78293088 & 1260 & 140 & 0.8252 & 27.44 & 7.2 \\
2450438.88091902 & 1800 & 140 & 0.3890 & 1.13  & 7.6 \\
2450440.72651416 & 1800 & 130 & 0.3368 & 6.08  & 5.9 \\
2450441.87574535 & 1260 & 140 & 0.9270 & 44.35 & 6.7 \\
2450446.65767851 & 2232 & 140 & 0.3828 & -0.36 & 6.2 \\
2450447.83059191 &  900 & 130 & 0.9851 & 43.29 & 8.0 \\
2450448.70624150 &  900 & 140 & 0.4348 & -6.85 & 6.8 \\
2450450.82337754 &  900 & 130 & 0.5221 & -7.65 & 7.6 \\
2450451.83619104 & 1260 & 120 & 0.0422 & 43.62 & 7.4 \\
2450453.74258670 & 1676 & 100 & 0.0212 & 43.78 & 7.8 \\
2450454.82479083 & 2700 & 120 & 0.5770 & -5.12 & 7.5 \\
2450456.83642216 & 3600 & 130 & 0.6101 & -1.62 & 7.2 \\
2450457.76155057 & 1440 & 130 & 0.0852 & 40.43 & 8.2 \\
\noalign{\smallskip}
\hline
\end{supertabular}
\vspace{3mm}


\mbox{KPNO, Mar 4 -- Mar 18, 1994}
\begin{supertabular}{ccccrc}
\hline
\noalign{\smallskip}
 HJD & $t\nm{exp}$ & S/N & phase & \multicolumn{1}{c}{$v_r$} & $\sigma_{v_r}$ \\
\noalign{\smallskip}
\hline
\noalign{\smallskip}
 2449416.61902162 & 1917 & 200 & 0.4071 & -2.40 & 3.0 \\
 2449417.61724752 & 3000 & 270 & 0.9197 & 43.65 & 4.4 \\
 2449426.60346990 & 3000 & 300 & 0.5346 & -7.75 & 3.4 \\
\noalign{\smallskip}
\hline
\end{supertabular}
\vspace{3mm}


\mbox{KPNO, Feb 22 -- Mar 7, 1995}
\begin{supertabular}{ccccrc}
\hline
\noalign{\smallskip}
 HJD & $t\nm{exp}$ & S/N & phase & \multicolumn{1}{c}{$v_r$} & $\sigma_{v_r}$ \\
\noalign{\smallskip}
\hline
\noalign{\smallskip}
 2449770.60061516 & 3000 & 200 & 0.1941 & 28.79 & 4.7 \\
 2449773.61609044 & 3000 & 200 & 0.7427 & 11.56 & 4.9 \\
 2449774.59440811 & 2554 & 200 & 0.2451 & 19.73 & 6.6 \\
 2449781.59376862 & 2204 & 180 & 0.8396 & 34.63 & 7.2 \\
 2449783.61495233 & 2312 & 210 & 0.8776 & 39.89 & 7.2 \\
\noalign{\smallskip}
\hline
\end{supertabular}
\vspace{3mm}


\mbox{KPNO, Jan 10 -- 24, 1996}
\begin{supertabular}{ccccrc}
\hline
\noalign{\smallskip}
 HJD & $t\nm{exp}$ & S/N & phase & \multicolumn{1}{c}{$v_r$} & $\sigma_{v_r}$ \\
\noalign{\smallskip}
\hline
\noalign{\smallskip}
 2450093.58895868 & 1800 & 310 & 0.0646 & 47.48 & 3.63 \\
 2450093.72198162 & 3600 & 320 & 0.1329 & 39.57 & 3.20 \\
 2450093.86983001 & 2429 & 330 & 0.2088 & 29.50 & 3.22 \\
 2450094.58586781 & 3600 & 270 & 0.5765 & -0.35 & 3.99 \\
 2450094.75631891 & 3600 & 190 & 0.6641 & 9.93  & 4.10 \\
 2450094.84253948 & 3600 & 280 & 0.7083 & 18.69 & 4.40 \\
 2450095.59933732 & 3600 & 350 & 0.0970 & 45.06 & 2.25 \\
 2450095.84999107 & 3600 & 360 & 0.2257 & 29.28 & 2.30 \\
 2450096.58167400 & 3600 & 370 & 0.6015 & 1.03  & 3.63 \\
 2450096.84057936 & 3600 & 360 & 0.7344 & 22.23 & 2.60 \\
 2450098.58923565 & 3676 & 300 & 0.6324 & 5.23  & 2.35 \\
 2450100.63909368 & 3600 & 240 & 0.6852 & 13.49 & 2.70 \\
 2450100.83270193 & 3600 & 310 & 0.7846 & 30.55 & 2.48 \\
 2450101.58188063 & 3600 & 220 & 0.1693 & 35.55 & 3.54 \\
 2450101.72781877 & 3600 & 330 & 0.2443 & 22.79 & 2.05 \\
 2450102.72229088 & 3600 & 270 & 0.7550 & 23.66 & 2.28 \\
 2450103.72935395 & 3600 & 190 & 0.2722 & 18.63 & 1.83 \\
 2450103.83519003 & 3600 & 210 & 0.3265 & 10.44 & 1.78 \\
 2450106.60196237 & 2705 & 180 & 0.7474 & 22.37 & 2.39 \\
 2450106.71444160 & 3600 & 250 & 0.8051 & 31.37 & 2.35 \\
 2450106.83906083 & 3005 & 350 & 0.8692 & 39.50 & 2.49 \\
 2450107.68453725 & 3600 & 190 & 0.3033 & 8.19  & 2.00 \\
 2450107.77017848 & 3600 & 280 & 0.3473 & 2.75  & 2.01 \\
 2450107.82899318 & 3600 & 220 & 0.3775 & 0.75  & 2.35 \\
\noalign{\smallskip}
\hline
\end{supertabular}
\vspace{3mm}


\mbox{KPNO, Dec 26, 1997 -- Jan 14, 1998.}
\begin{supertabular}{ccccrc}
\hline
\noalign{\smallskip}
 HJD & $t\nm{exp}$ & S/N & phase & \multicolumn{1}{c}{$v_r$} & $\sigma_{v_r}$ \\
\noalign{\smallskip}
\hline
\noalign{\smallskip}
 2450809.58273645 & 3640 & 299 & 0.7627 & 20.68 & 2.2 \\
 2450809.72859092 & 5135 & 294 & 0.8376 & 32.50 & 1.9 \\
 2450809.86952719 & 5078 & 237 & 0.9100 & 42.79 & 2.0 \\
 2450810.58576035 & 3642 & 288 & 0.2778 & 14.96 & 2.0 \\
 2450810.68774591 & 4798 & 221 & 0.3302 & 6.80  & 2.2 \\
 2450810.82391309 & 4555 & 153 & 0.4001 & -0.53 & 2.0 \\
 2450811.62176909 & 3672 & 188 & 0.8099 & 28.22 & 1.8 \\
 2450811.72285058 & 3642 & 198 & 0.8618 & 36.34 & 2.3 \\
 2450811.82025871 & 3639 & 232 & 0.9118 & 42.17 & 1.9 \\
 2450811.90381865 & 2727 & 201 & 0.9547 & 45.86 & 1.6 \\
 2450812.86269379 &  900 & 81  & 0.4471 & -5.48 & 1.9 \\
 2450813.58682342 & 3643 & 312 & 0.8190 & 30.62 & 2.3 \\
 2450813.68721789 & 3642 & 372 & 0.8706 & 37.31 & 1.9 \\
 2450813.80771387 & 8085 & 389 & 0.9325 & 45.17 & 1.5 \\
 2450813.89197395 & 3642 & 333 & 0.9757 & 47.30 & 1.8 \\
 2450814.58051026 & 3640 & 348 & 0.3293 & 7.42  & 1.7 \\
 2450814.66755949 & 3641 & 383 & 0.3740 & 1.37  & 2.2 \\
 2450814.72819247 & 3445 & 376 & 0.4052 & -1.22 & 1.9 \\
 2450814.81155554 & 3241 & 368 & 0.4480 & -3.79 & 2.3 \\
 2450814.91346455 & 2300 & 107 & 0.5003 & -8.52 & 2.1 \\
 2450815.58646849 & 3544 & 360 & 0.8459 & 34.89 & 2.2 \\
 2450815.68379471 & 3641 & 381 & 0.8959 & 40.47 & 2.2 \\
 2450815.79300152 & 3643 & 361 & 0.9520 & 44.52 & 2.2 \\
 2450815.85301531 &  900 & 184 & 0.9828 & 48.18 & 1.9 \\
 2450816.58303964 & 3642 & 319 & 0.3577 & 2.53  & 1.8 \\
 2450819.66147317 & 3643 & 313 & 0.9387 & 43.50 & 2.1 \\
 2450819.77794707 & 3641 & 323 & 0.9985 & 46.63 & 2.0 \\
 2450819.87733875 & 3641 & 280 & 0.0495 & 44.08 & 2.2 \\
 2450820.57320682 & 3643 & 328 & 0.4069 & -3.37 & 1.7 \\
 2450820.68061358 & 3644 & 302 & 0.4620 & -5.41 & 1.7 \\
 2450820.78112776 & 3642 & 334 & 0.5137 & -6.18 & 1.6 \\
 2450820.87562960 & 3643 & 265 & 0.5622 & -4.86 & 1.8 \\
 2450821.57235406 & 3643 & 332 & 0.9200 & 43.51 & 1.7 \\
 2450821.67462729 & 3643 & 361 & 0.9725 & 47.03 & 1.5 \\
 2450821.77773380 & 3643 & 363 & 0.0255 & 47.53 & 1.7 \\
 2450821.88964179 & 1830 & 91  & 0.0829 & 42.76 & 1.8 \\
 2450822.71209769 & 4820 & 308 & 0.5053 & -7.25 & 2.0 \\
 2450822.81245542 & 3643 & 368 & 0.5568 & -4.62 & 1.8 \\
 2450822.86863216 & 3641 & 325 & 0.5857 & -2.93 & 2.1 \\
 2450823.78021880 & 3644 & 362 & 0.0538 & 45.33 & 2.0 \\
 2450825.62981148 & 3644 & 368 & 0.0037 & 46.19 & 1.9 \\
 2450825.71605559 & 3642 & 361 & 0.0480 & 44.99 & 1.8 \\
 2450825.78051100 & 3641 & 341 & 0.0811 & 43.04 & 2.1 \\
 2450825.86746150 & 3643 & 253 & 0.1257 & 38.20 & 1.6 \\
 2450826.58727787 & 3642 & 358 & 0.4954 & -7.26 & 2.0 \\
 2450826.71078173 & 3641 & 374 & 0.5588 & -5.73 & 2.1 \\
 2450826.77149497 & 3641 & 356 & 0.5900 & -3.41 & 1.8 \\
 2450826.87359944 & 2728 & 124 & 0.6424 & 2.18  & 2.2 \\
 2450827.70428623 & 3653 & 163 & 0.0690 & 43.71 & 2.2 \\
 2450827.77446631 & 3653 & 342 & 0.1051 & 41.39 & 2.0 \\
 2450827.84209449 & 3644 & 151 & 0.1398 & 37.29 & 2.1 \\
 2450827.87402503 & 3674 & 122 & 0.1562 & 35.47 & 2.0 \\
 2450828.57683812 & 3643 & 158 & 0.5171 & -6.92 & 2.0 \\
 2450828.59804605 & 3650 & 330 & 0.5280 & -7.01 & 2.0 \\
 2450828.70171256 & 3643 & 310 & 0.5813 & -4.48 & 1.7 \\
 2450828.82504772 & 3645 & 301 & 0.6446 & 1.91  & 2.3 \\
 2450828.86755618 & 3647 & 280 & 0.6664 & 4.78  & 1.9 \\
\noalign{\smallskip}
\hline
\end{supertabular}
\vspace{4mm}

%
%
%
%
%

\mbox{MUSICOS, Nov 22, 1998 -- Dec 13, 1998.}
\begin{supertabular}{@{}rccccr@{\,$\pm$\,}l@{}}
\hline
\hline
Site & HJD & $t\nm{exp}$ & S/N & phase & \multicolumn{2}{c}{$v_r$} \\
 & $[$24511+$]$ & $[$s$]$ & & & \multicolumn{2}{c}{$[$\kms$]$} \\
\hline
\hline

OHP193 & 41.37584 & 1200 & 200 & 0.1549 & 35.1 & 3.8 \\
OHP193 & 41.48556 & 1200 & 190 & 0.2112 & 26.8 & 3.1 \\
OHP193 & 41.60501 & 1200 & 180 & 0.2726 & 23.4 & 4.4 \\
OHP193 & 44.37304 & 1800 & 170 & 0.6941 & 9.5  & 3.1 \\
OHP193 & 45.37442 & 1800 & 100 & 0.2083 & 27.3 & 4.8 \\
OHP193 & 46.36989 & 2100 & 210 & 0.7196 & 13.2 & 3.6 \\
OHP193 & 47.45078 & 2400 & 180 & 0.2747 & 16.2 & 3.0 \\

\hline

SAAO & (22:09) & 1800 & (80) & -- & \multicolumn{2}{c}{--} \\
SAAO & (21:29) & 924  &  & -- & \multicolumn{2}{c}{--} \\
SAAO & (21:13) & 1800 & (90) & -- & \multicolumn{2}{c}{--} \\
SAAO & (00:52) & 1800 &  & -- & \multicolumn{2}{c}{--} \\

\hline

BXO & 49.25642 & 5400 &  30 & 0.2019 & -6.4 & 2.8 \\
BXO & 50.10649 & 3000 &  80 & 0.6385 & 61.8 & 1.2 \\
BXO & 50.21550 & 3000 & 140 & 0.6945 & 69.9 & 1.7 \\
BXO & 50.27836 & 3000 & 150 & 0.7268 & 75.3 & 1.4 \\
BXO & 52.09376 & 3000 & 120 & 0.6591 & 3.2  & 2.7 \\
BXO & 52.19891 & 3000 & 120 & 0.7131 & 12.3 & 3.3 \\
BXO & 52.27990 & 3000 &  80 & 0.7547 & 18.3 & 2.5 \\
BXO & 55.15105 & 2700 &  30 & 0.2291 & 49.7 & 2.3 \\
BXO & 55.24648 & 3000 &  40 & 0.2781 & 48.6 & 1.7 \\

\hline

MSO & 47.04042 & 1500 & 95  & 0.0639 & 49.3 & 4.6 \\
MSO & 47.05919 & 1500 & 120 & 0.0736 & 49.7 & 3.1 \\
MSO & 47.16410 & 1500 & 90  & 0.1274 & 43.3 & 2.2 \\
MSO & 47.94040 & 1500 & 100 & 0.5261 & 0.6  & 4.5 \\
MSO & 48.02403 & 1500 & 130 & 0.5691 & 0.9  & 1.9 \\
MSO & 48.15425 & 1200 & 100 & 0.6359 & 6.9  & 2.3 \\
MSO & 48.93913 & 1500 & 110 & 0.0390 & 51.3 & 3.6 \\
MSO & 49.07460 & 1500 & 110 & 0.1086 & 45.7 & 3.7 \\
MSO & 49.15082 & 1200 & 90  & 0.1477 & 39.8 & 6.2 \\
MSO & 49.94285 & 1500 & 25  & 0.5545 & -4.0 & 6.4 \\
MSO & 50.98270 & 1500 & 105 & 0.0885 & 49.3 & 3.3 \\
MSO & 51.04591 & 1200 & 110 & 0.1209 & 44.8 & 2.6 \\
MSO & 51.10611 & 1200 & 110 & 0.1519 & 41.8 & 3.6 \\

\hline

KPNO & 41.63019 & 1200 & 45 & 0.2855 & \multicolumn{2}{c}{--} \\
KPNO & 41.67969 & 1200 & 50 & 0.3109 & \multicolumn{2}{c}{--} \\
KPNO & 41.78911 & 1200 & 40 & 0.3671 & \multicolumn{2}{c}{--} \\
KPNO & 41.92587 & 1200 & 70 & 0.4373 & \multicolumn{2}{c}{--} \\
KPNO & 42.61928 & 1200 & 55 & 0.7934 & \multicolumn{2}{c}{--} \\
KPNO & 42.70574 & 1200 & 70 & 0.8378 & \multicolumn{2}{c}{--} \\
KPNO & 42.80249 & 1200 & 60 & 0.8875 & \multicolumn{2}{c}{--} \\
KPNO & 42.90963 & 1200 & 65 & 0.9426 & \multicolumn{2}{c}{--} \\
KPNO & 42.98487 & 1200 & 65 & 0.9812 & \multicolumn{2}{c}{--} \\
KPNO & 43.80302 & 1500 & 80 & 0.4014 & \multicolumn{2}{c}{--} \\
KPNO & 43.88231 & 1200 & 70 & 0.4421 & \multicolumn{2}{c}{--} \\
KPNO & 43.89726 & 1200 & 70 & 0.4498 & \multicolumn{2}{c}{--} \\
KPNO & 43.91224 & 1200 & 70 & 0.4574 & \multicolumn{2}{c}{--} \\
KPNO & 44.70456 & 1500 & 80 & 0.8643 & \multicolumn{2}{c}{--} \\
KPNO & 44.72323 & 1500 & 80 & 0.8739 & \multicolumn{2}{c}{--} \\
KPNO & 44.74190 & 1500 & 80 & 0.8835 & \multicolumn{2}{c}{--} \\
KPNO & 44.86733 & 1500 & 80 & 0.9479 & \multicolumn{2}{c}{--} \\
KPNO & 44.88594 & 1500 & 70 & 0.9575 & \multicolumn{2}{c}{--} \\
KPNO & 44.90453 & 1500 & 65 & 0.9670 & \multicolumn{2}{c}{--} \\

\hline

ESO90 & 46.58071 & 2700 & 170 & 0.8278 & \multicolumn{2}{c}{--} \\
ESO90 & 47.56183 & 2700 & 170 & 0.3317 & \multicolumn{2}{c}{--} \\
ESO90 & 49.56834 & 2700 & 160 & 0.3621 & \multicolumn{2}{c}{--} \\
ESO90 & 50.56477 & 2700 & 200 & 0.8738 & \multicolumn{2}{c}{--} \\
ESO90 & 51.56331 & 2700 & 170 & 0.3866 & \multicolumn{2}{c}{--} \\
ESO90 & 52.56155 & 2700 & 250 & 0.8993 & \multicolumn{2}{c}{--} \\
ESO90 & 53.59846 & 2700 & 130 & 0.4318 & \multicolumn{2}{c}{--} \\
ESO90 & 54.59340 & 2700 & 150 & 0.9427 & \multicolumn{2}{c}{--} \\
ESO90 & 55.59898 & 2700 & 170 & 0.4592 & \multicolumn{2}{c}{--} \\
ESO90 & 56.59905 & 2700 & 170 & 0.9727 & \multicolumn{2}{c}{--} \\
ESO90 & 57.59908 & 2700 & 200 & 0.4863 & \multicolumn{2}{c}{--} \\
ESO90 & 58.59838 & 2700 & 170 & 0.9995 & \multicolumn{2}{c}{--} \\
ESO90 & 59.59979 & 2700 & 170 & 0.5138 & \multicolumn{2}{c}{--} \\
ESO90 & 60.63132 & 2700 & 170 & 0.0435 & \multicolumn{2}{c}{--} \\

\hline

ESO152 & 43.74576 & 600 & 270 & 0.3720 & \multicolumn{2}{c}{--} \\
ESO152 & 43.85290 & 600 & 190 & 0.4270 & \multicolumn{2}{c}{--} \\
ESO152 & 44.59754 & 600 & 280 & 0.8094 & \multicolumn{2}{c}{--} \\
ESO152 & 44.85283 & 546 & 170 & 0.9405 & \multicolumn{2}{c}{--} \\
ESO152 & 45.60065 & 600 & 290 & 0.3245 & \multicolumn{2}{c}{--} \\
ESO152 & 45.80423 & 600 & 270 & 0.4291 & \multicolumn{2}{c}{--} \\
ESO152 & 46.74656 & 600 & 270 & 0.9130 & \multicolumn{2}{c}{--} \\
ESO152 & 46.83205 & 600 & 310 & 0.9569 & \multicolumn{2}{c}{--} \\
ESO152 & 47.72803 & 600 & 330 & 0.4170 & \multicolumn{2}{c}{--} \\
ESO152 & 47.81225 & 600 & 320 & 0.4603 & \multicolumn{2}{c}{--} \\
ESO152 & 48.67180 & 600 & 350 & 0.9017 & \multicolumn{2}{c}{--} \\
ESO152 & 48.76177 & 600 & 280 & 0.9479 & \multicolumn{2}{c}{--} \\
ESO152 & 49.70264 & 600 & 250 & 0.4311 & \multicolumn{2}{c}{--} \\
ESO152 & 49.78489 & 600 & 290 & 0.4733 & \multicolumn{2}{c}{--} \\
ESO152 & 50.66693 & 600 & 270 & 0.9263 & \multicolumn{2}{c}{--} \\
ESO152 & 50.74739 & 600 & 280 & 0.9676 & \multicolumn{2}{c}{--} \\
ESO152 & 51.71147 & 600 & 300 & 0.4627 & \multicolumn{2}{c}{--} \\
ESO152 & 51.79690 & 600 & 220 & 0.5066 & \multicolumn{2}{c}{--} \\

\hline

INT & 52.50107 & 1500 &  70 & 0.8682 & \multicolumn{2}{c}{--} \\
INT & 54.53835 & 1500 &  50 & 0.9145 & \multicolumn{2}{c}{--} \\
INT & 55.38794 & 1500 &  50 & 0.3508 & \multicolumn{2}{c}{--} \\
INT & 55.64474 & 1500 &  90 & 0.4827 & \multicolumn{2}{c}{--} \\
INT & 56.36766 & 1500 &  90 & 0.8539 & \multicolumn{2}{c}{--} \\
INT & 56.43590 & 1500 & 110 & 0.8890 & \multicolumn{2}{c}{--} \\
INT & 56.63331 & 1500 & 110 & 0.9903 & \multicolumn{2}{c}{--} \\
INT & 57.39970 & 1200 & 160 & 0.3839 & \multicolumn{2}{c}{--} \\
INT & 57.47034 & 1500 & 160 & 0.4202 & \multicolumn{2}{c}{--} \\
INT & 57.64745 & 1500 & 170 & 0.5112 & \multicolumn{2}{c}{--} \\
INT & 58.53580 & 1500 & 120 & 0.9674 & \multicolumn{2}{c}{--} \\

\hline

OHP152 & 39.42100 & 1200 &  & 0.1510 & \multicolumn{2}{c}{--} \\
OHP152 & 40.39846 & 3000 &  & 0.6529 & \multicolumn{2}{c}{--} \\
OHP152 & 40.50785 & 3000 &  & 0.7091 & \multicolumn{2}{c}{--} \\
OHP152 & 40.60826 & 3000 &  & 0.7607 & \multicolumn{2}{c}{--} \\
OHP152 & 43.55433 & 3000 &  & 0.2736 & \multicolumn{2}{c}{--} \\
OHP152 & 44.37714 & 3000 &  & 0.6962 & \multicolumn{2}{c}{--} \\
OHP152 & 45.38122 & 3600 &  & 0.2118 & \multicolumn{2}{c}{--} \\
OHP152 & 45.49359 & 3000 &  & 0.2695 & \multicolumn{2}{c}{--} \\
OHP152 &  45.5876 & 3000 &  & 0.3178 & \multicolumn{2}{c}{--} \\
OHP152 & 46.38628 & 2292 &  & 0.7280 & \multicolumn{2}{c}{--} \\
OHP152 & 47.48551 & 2991 &  & 0.2925 & \multicolumn{2}{c}{--} \\
OHP152 & 47.49585 & 3000 &  & 0.2978 & \multicolumn{2}{c}{--} \\
OHP152 & 47.58590 & 3600 &  & 0.3441 & \multicolumn{2}{c}{--} \\

\hline
\end{supertabular} 
\vspace{4mm}

%
%
%
%
%

Additional radial velocities from the literature. The values were, if possible, corrected
to match the RV standard star values of \citet{scarfe90}.
\begin{supertabular}{ccr@{\,}c@{}c}
\hline
\noalign{\smallskip}
 HJD & phase & \multicolumn{1}{c}{$v_r$} & Instrument & Ref. \\
$[$244+$]$ & & $[$\kms$]$ & & \\
\noalign{\smallskip}
\hline
\noalign{\smallskip}
 9346.1467 & 0.2161 & $24.5 \pm 1$ & AAT/UCLES & D \\
 9348.1099 & 0.2243 & $23.5 \pm 1$ & AAT/UCLES & D \\
\noalign{\smallskip}
\hline
\noalign{\smallskip}
 9643.9022 & 0.1281 & $28.5 \pm  6.9$ & Lowell/JH\,1.1m/SSS & G \\
 9645.9176 & 0.1634 & $45.5 \pm  2.8$ & Lowell/JH\,1.1m/SSS & G \\
 9647.8882 & 0.1751 & $30.9 \pm  3.6$ & Lowell/JH\,1.1m/SSS & G \\
 9651.8605 & 0.2154 & $19.0 \pm 10.1$ & Lowell/JH\,1.1m/SSS & G \\
 9655.8660 & 0.2722 & $14.4 \pm  2.3$ & Lowell/JH\,1.1m/SSS & G \\
\noalign{\smallskip}
\hline
\noalign{\smallskip}
 7150.6665 & 0.7286 &  13.03 & KPNO/CF & S \\
 7150.7295 & 0.7609 &  19.23 & KPNO/CF & S \\
 7150.7964 & 0.7953 &  24.93 & KPNO/CF & S \\
 7150.9253 & 0.8615 &  32.63 & KPNO/CF & S \\
 7151.5811 & 0.1983 &  27.30 & KPNO/CF & S \\
 7151.7505 & 0.2852 &  14.19 & KPNO/CF & S \\
 7151.8926 & 0.3582 &   2.45 & KPNO/CF & S \\
 7152.5664 & 0.7043 &   9.99 & KPNO/CF & S \\
 7152.7280 & 0.7872 &  22.95 & KPNO/CF & S \\
 7156.9126 & 0.9362 &  42.33 & KPNO/CF & S \\
 7157.5918 & 0.2850 &  12.49 & KPNO/CF & S \\
 7157.6611 & 0.3206 &   5.49 & KPNO/CF & S \\
 7157.8130 & 0.3986 &  -3.21 & KPNO/CF & S \\
 7157.9033 & 0.4450 &  -6.47 & KPNO/CF & S \\
 7163.6997 & 0.4218 & -10.2  & KPNO/CF & S \\
 7163.7451 & 0.4451 &  -6.0  & KPNO/CF & S \\
 7163.8130 & 0.4799 & -10.87 & KPNO/CF & S \\
 7163.8486 & 0.4982 & -10.87 & KPNO/CF & S \\
 7163.8882 & 0.5186 & -11.07 & KPNO/CF & S \\
 7170.7617 & 0.0484 &  43.0  & KPNO/CF & S \\
 7170.8936 & 0.1162 &  38.5  & KPNO/CF & S \\
\noalign{\smallskip}
\hline
\noalign{\smallskip}
 4179.903 & 0.0948 &  43.15 & McDonald 2.7m & F \\
 4180.916 & 0.6150 &  -3.77 & McDonald 2.1m & F \\
 4181.857 & 0.0983 &  40.79 & McDonald 2.1m & F \\
 4182.814 & 0.5897 &  -8.81 & McDonald 2.1m & F \\
 4473.973 & 0.1143 &  38.15 & McDonald 2.7m & F \\
 4475.963 & 0.1362 &  35.65 & McDonald 2.7m & F \\
 4478.955 & 0.6728 &   4.09 & McDonald 2.1m & F \\
 4507.885 & 0.5298 &  -9.2  & KPNO          & F \\
 4627.626 & 0.0227 &  44.43 & McDonald 2.7m & F \\
 5356.702 & 0.4392 &  -6.37 & KPNO/CF/RCA   & F \\
 5357.670 & 0.9363 &  41.93 & KPNO/CF/RCA   & F \\
 5358.679 & 0.4545 & -10.41 & KPNO/CF/RCA   & F \\
 5359.745 & 0.0019 &  47.59 & KPNO/CF/RCA   & F \\
 5360.716 & 0.5006 &  -7.07 & KPNO/CF/RCA   & F \\
 5361.702 & 0.0069 &  41.23 & KPNO/CF/RCA   & F \\
 5594.999 & 0.8165 &  27.89 & KPNO/CF/TI    & F \\
 5596.022 & 0.3418 &   3.13 & KPNO/CF/TI    & F \\
 5596.963 & 0.8251 &  27.63 & KPNO/CF/TI    & F \\
 5599.008 & 0.8753 &  36.25 & KPNO/CF/TI    & F \\
 5717.791 & 0.8762 &  34.75 & KPNO/CF/TI    & F \\
 5718.764 & 0.3759 &   1.83 & KPNO/CF/TI    & F \\
 5719.768 & 0.8915 &  39.83 & KPNO/CF/TI    & F \\
 5720.740 & 0.3907 &  -3.75 & KPNO/CF/TI    & F \\
 5721.727 & 0.8976 &  40.63 & KPNO/CF/TI    & F \\
 5941.958 & 0.9971 &  45.3  & KPNO/CF/TI    & F \\
 5971.996 & 0.4231 &  -7.11 & KPNO/CF/TI    & F \\
 6076.766 & 0.2276 &  21.99 & KPNO/CF/TI    & F \\
\noalign{\smallskip}
\hline
\noalign{\medskip}
\multicolumn{5}{l}{Reference code:}\\
& \multicolumn{4}{l}{D \dots\ \citet{donati:semel97}} \\
& \multicolumn{4}{l}{F \dots\ \citet{fekel:quigley87}} \\
& \multicolumn{4}{l}{G \dots\ \citet{gunn:hall96}} \\
& \multicolumn{4}{l}{S \dots\ \citet{kgs90}} \\
\end{supertabular}

\end{center}



\end{document}